\documentclass[%
    amsmath,
    amssymb,
    aps,
    superscriptaddress,
    footinbib,
    prx,
    10pt
]{revtex4-2}

\usepackage{customarticlestyle}
\usepackage[capitalise]{cleveref}
\usepackage{tikz}
\usepackage{tikz_options}

\begin{document}

\title{Stabilizer-based quantum simulation of fermion dynamics with local qubit encodings}

\author{Anthony Gandon$^{\orcidanthony}$}
\affiliation{\ibmquantum}
\affiliation{\ethzurich}
\author{Samuele Piccinelli}
\affiliation{\ibmquantum}
\affiliation{\EPFL}
\author{Max Rossmannek$^{\orcidmax}$}
\affiliation{\ibmquantum}
\author{Francesco Tacchino$^{\orcidfrancesco}$}
\affiliation{\ibmquantum}
\author{Alberto Baiardi}
\affiliation{\ibmquantum}
\author{Jannes Nys$^{\orcidjannes}$}
\affiliation{\ethzurich}
\author{Ivano Tavernelli$^{\orcidivano}$}
\affiliation{\ibmquantum}

\begin{abstract}
Simulating the dynamical properties of large-scale many-fermion systems is a longstanding goal of quantum chemistry, material science and condensed matter. Local fermion-to-qubit encodings have opened a new path for practical fermionic simulations on digital quantum hardware where fermionic statistics are not enforced at the hardware level. In this paper, we explore these local encodings from the perspective of the corresponding time-evolution unitaries. Specifically, we propose a new framework for digital implementations of these qubit-encoded fermionic time-evolution unitaries based on \emph{flow sets}, which are one-dimensional subsets of the directed fermionic interaction graph. We find that any local fermionic encoding, when restricted to a given flow set, adopts a simple structure that we can classify systematically. For each categorized flow-set form, we propose a low-depth qubit quantum circuit that implements the time evolution unitary using the stabilizer formalism.
As an application of our construction, we introduce novel flow-based decompositions for known two-dimensional encodings, leading to efficient circuit decompositions of time-evolution unitaries. We generally observe a space-time trade-off, where mappings with larger qubit-to-fermion ratios yield shallower time-evolution quantum circuits.
\end{abstract}

\maketitle

\section{Introduction}

Accurate descriptions of complex many-body systems often rely on the microscopic description of their constituents, which, for quantum-chemistry and material-science applications, are described by fermionic statistics. In this work, we consider fermionic systems in second quantization, \textit{i.e.}, described through a finite basis set (usually defined as \textit{fermionic modes}), and focus on lattice Hamiltonians with spatially local interactions.

Recent advances in the engineering and control of quantum degrees of freedom in various experimental platforms have opened new avenues for simulating the real-time dynamics of quantum many-body systems~\citep{Feynman1982SimulatingPhysicsComputers,Jordan2018BQPcompletenessScatteringScalar,Clinton2024NeartermQuantumSimulationa}. While fermionic quantum processing~\citep{Bravyi2002FermionicQuantumComputationa,Schuckert2024FermionqubitFaulttolerantQuantum} has already been demonstrated in neutral atom arrays~\citep{Gonzalez-Cuadra2023FermionicQuantumProcessing}, the fermionic statistics are not required to achieve universal quantum computation~\citep{Tacchino2020QuantumComputersUniversal}. Prominent platforms for digital quantum computing instead use collections of distinguishable local quantum systems such as qubits. In this context, fermionic statistics are not enforced by the hardware itself, but instead emerge from a suitable set of qubit operators~\citep{Ball2005FermionsFermionFields}.

Fermionic encodings establish equivalences between fermionic and qubit (or bosonic) operator algebras. The seminal example is the Jordan-Wigner equivalence between a one-dimensional lattice of $N_f$ fermionic modes and a one-dimensional qubit chain with $N_q=N_f$ qubits. For fermionic lattices in dimension $d>2$, the requirement that local fermionic operators get mapped to purely local operators on the target qubit lattice rules out any extension to one-dimensional ordering of the $d$-dimensional lattice sites~\citep{Chen2018ExactBosonizationTwoa,Tantivasadakarn2020JordanWignerDualitiesTranslationinvariant}. Instead, locality can be recovered at the cost of increasing the number of qubit degrees of freedom in the target space ($N_q>N_f$)~\citep{Wosiek1982LocalRepresentationFermions,Verstraete2005MappingLocalHamiltoniansa,Bochniak2020BosonizationBasedClifford,Derby2021CompactFermionQubit,Verstraete2005MappingLocalHamiltoniansa,Steudtner2019QuantumCodesQuantum,Derby2021CompactFermionQubita} and possibly long-range entangled states~\cite{Guaita2025LocalityQubitEncodings}. The main difference with the one-dimensional case is that higher-dimensional fermionic systems are dual to qubit systems on extended lattices and constrained by local commuting symmetries~\citep{Chen2018ExactBosonizationTwoa,Chen2023EquivalenceFermiontoQubitMappings,Tantivasadakarn2020JordanWignerDualitiesTranslationinvariant,Levin2003FermionsStringsGauge}. Such symmetries have arbitrary code distances~\citep{Algaba2025FermiontoqubitEncodingsArbitrary}, and can be used to detect and correct errors in the framework of quantum error correction. These works serve as the foundation for simulations of fermionic lattice Hamiltonians with qubit-based quantum hardware.

More recently, Chen \emph{et al.}~\citep{Chen2023EquivalenceFermiontoQubitMappings} showed that locality-preserving encodings of fermionic operators on a two-dimensional lattice are all equivalent up to Clifford unitary transformations. However, in practice, two-dimensional fermionic encodings differ in the qubit-to-fermion ratios, gauge constraints, and locality of the encoded qubit operators. So far, the complexity of the resulting unitary time-evolution quantum circuits, after compilation and optimization, for applications in digital quantum computing has been studied independently for each specific fermionic encoding~\citep{Campbell2022EarlyFaulttolerantSimulations,Nigmatullin2025ExperimentalDemonstrationBreakeven,Dyrenkova2025ScalableSimulationFermionic}.

This work presents a unified strategy for comparing the complexity of fermionic time-evolution unitaries when implemented on qubit-based processors that support the universal gate set of quantum computing. Instead of compiling the qubit-encoded unitary after the fermionic mapping, our proposed approach is first formulated on the fermionic time-evolution unitary. More precisely, we introduce a systematic grouping of Hamiltonian terms into Majorana stabilizer groups~\citep{Mudassar2024EncodingMajoranaCodes,Bettaque2025StructureMajoranaClifford}, which are sets of commuting fermionic hoppings, and which we will refer to as \emph{flow sets}. These groups of Majorana operators can be applied in parallel, independently of the qubit encoding, by exploiting the well-established toolbox of stabilizer states and error correction~\citep{Gottesman1996ClassQuantumErrorcorrecting,Gottesman1997StabilizerCodesQuantum} in the context of Trotterized time-evolution unitaries for fermions.
Although the previous strategy is presented at fermionic level, we show that it also generally applies to qubit representations of the fermionic operators under some elementary conditions. The Majorana stabilizer group~\citep{Mudassar2024EncodingMajoranaCodes,Bettaque2025StructureMajoranaClifford} corresponding to a flow set becomes a Pauli stabilizer group~\citep{Gottesman1996ClassQuantumErrorcorrecting,Gottesman1997StabilizerCodesQuantum} after the fermion-to-qubit mapping. The implementation of the qubit unitary representation of the fermionic evolution is then limited by the depth of the Clifford stabilizer-encoding circuit~\citep{Higgott2021OptimalLocalUnitary,Berg2020CircuitOptimizationHamiltonian}, which becomes the relevant metric for digital quantum computing applications.

This paper is organized as follows. In \cref{sec:fermionic_systems}, we describe fermionic systems and summarize key properties of fermionic operators. In \cref{sec:trotterization} we introduce the minimal example of a kinetic lattice Hamiltonian that will serve as an illustration of our method. We also define \emph{flow sets} to generalize the traditional operator grouping approach for the Trotterization and parallelization of fermionic hopping operators.
We find that any local fermionic encoding, when restricted to a given flow set, takes a simple one-dimensional form that we categorize systematically in~\cref{sec:qubit_representation_fermionic_operators}. For each categorized structure, we propose in \cref{sec:encoding_unitaries_1D} a stabilizer-based qubit unitary circuit to implement the time evolution under the encoded qubit operators. These, in turn, are used in \cref{sec:clifford_pauli_encoding} to reveal new efficient decompositions of fermionic time-evolution unitaries for common mappings in the literature. These decompositions heavily rely on the generalized flow set definitions for parallelizing the Hamiltonian's hopping terms, and are less dependent on the Pauli weight of the encoded hopping operators.

\section{Fermionic operators and algebra} \label{sec:fermionic_systems}
Fermions are characterized by antisymmetric exchange statistics. In second quantization, for a system with $N_f$ fermionic modes, the creation and annihilation operators $c_j^\dag$, $c_k$ representing the creation (\emph{resp.} annihilation) of a fermionic particle on the modes $j,~k$ must satisfy the following anti-commutation relations
\begin{align}\label{eq:fermionic_canonical_relations}
    \{c_j^\dag, c_k\} = \delta_{jk},~~\{c_j, c_k\} = 0,~~ \{c_j^\dag, c_k^\dag\} = 0
\end{align}
regardless of the indices $j$ and $k$.
When representing the creation and annihilation operators on collections of local quantum systems such as qubits, the anticommutation relations generally result in non-local operators.
For convenience, we also introduce the $2 N_f$ Majorana operators associated with these fermionic modes $\gamma_{2j-1} = c_j + c_j^\dag,~ \gamma_{2j} = -i(c_j - c_j^\dag)$ which satisfy unified anti-commutation relations
\begin{align}\label{eq:majorana_canonical_relations}
    \{\gamma_a, \gamma_b\} = 2 \delta_{ab} I\, .
\end{align}

\begin{figure}[!ht]
    \centering
    \input{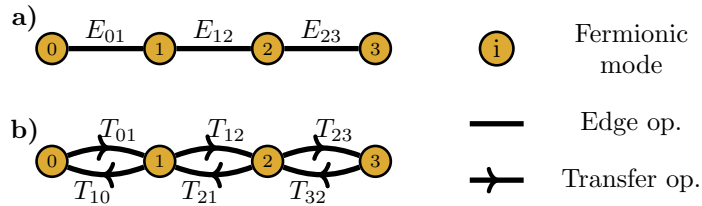}
    \caption{Graphical notation for representing the mixed-commutation relations of a one-dimensional fermionic system with $N_f=4$ sites. \textbf{a)} Edge fermionic operators are represented by undirected edges of the system's interaction graph, with overlap of two edges at a node implying the anti-commutation of the corresponding edge operators. Equivalently, edge operators can be represented as double-headed arrows to match the flow/clash property of the transfer operators. \textbf{b)} Transfer fermionic operators are represented as directed edges. The overlap of two edges at a node now implies anti-commutation relations if the edges are both directed either toward or from the node, \emph{i.e.} if they \emph{clash}, and commutation if there is a \emph{flow} of the arrows through the node. For example, the transfer operators $T_{01}$ and $T_{12}$ commute while $T_{01}$ and $T_{21}$ anticommute.}
    \label{fig:graphical_notation_p1}
\end{figure}

The parity superselection rule \citep{Wick1952IntrinsicParityElementary,Vidal2021QuantumOperationsInformation} imposes a fundamental constraint on fermionic states by forbidding superpositions of even and odd fermion parity states. For this reason, we will restrict ourselves to the algebra of observables with even fermionic parity, which is generated by the quadratic fermion operators $c_jc_k$, $c_j^\dag c_k$, $c_j c_k^\dag$, and $c_j^\dag c_k^\dag$. Hermitian operators within this algebra can be more conveniently expressed in terms of Hermitian generators constructed from the Majorana operators. One common choice for these generators is the edge-vertex ($E_{jk}, V_j$) operators 
\begin{align}\label{e:transfer_vertex_operators}
    V_j &= - i \gamma_{2j-1} \gamma_{2j} = -(c_j c_j - c_jc_j^\dag + c_j^\dag c_j - c_j^\dag c_j^\dag)= 1 - 2 c_j^\dag c_j \, ,\\
    E_{jk} &= - i \gamma_{2j-1} \gamma_{2k-1} = - i(c_jc_k + c_jc_k^\dag + c_j^\dag c_k + c_j^\dag c_k^\dag) = -E_{kj}\, .
\end{align}
which can be shown to satisfy the mixed fermionic-bosonic commutation relations for $j\neq k \neq l \neq m$
\begin{alignat}{3} \label{eq:mixed_edge_vertex_commutation_relations}
    & \textcolor{white}{[V_k, V_l]=0}  \quad  \{E_{jk}, V_k\}&&= 0  \quad \{E_{jk}, E_{kl}\} &&= 0 \notag  \\
    & [V_k, V_l]=0   \quad  \,\,  [E_{jk}, V_l]&&= 0  \quad   [E_{jk}, E_{lm}]&&= 0\, .
\end{alignat}
In the mixed relations in \cref{eq:mixed_edge_vertex_commutation_relations}, edge and vertex operators commute, unless they share an index, in which case they anticommute. This definition indicates that, when considering even-parity operators, the locality of the operator can be partially recovered since two space-separated quadratic operators commute. For convenience, we also introduce the so-called transfer operators ($T_{jk}$) defined from the relation
\begin{equation}
    T_{jk} = \frac{i}{2} V_j E_{jk} = \frac{i}{2} \gamma_{2j} \gamma_{2k-1}, \quad T_{kj} = \frac{i}{2} E_{jk} V_k = -\frac{i}{2} \gamma_{2j-1} \gamma_{2k}\, .
\end{equation}
The equality $E_{jk} = - E_{kj}$ for edge-generators translates to a product equality of the form ${T_{jk}=-V_jV_k T_{kj}}$. Transfer- and vertex- operators also generate the full Hermitian even-parity algebra but satisfy altered mixed commutation relations for $j<k<l<m$
\begin{equation} \label{eq:mixed_tranfer_vertex_commutation_relations}
\begin{matrix}
    & \{T_{jk}, V_k\} = 0 & \{T_{jk}, T_{lk}\} = 0 \, , & \\
    [V_k, V_l] = 0 & [T_{jk}, V_l] = 0 & [T_{jk}, T_{lm}] = 0 & [T_{jk}, T_{kj}] = 0 & [T_{jk}, T_{km}] = 0 \, .
\end{matrix}
\end{equation}
To facilitate the manipulation of the mixed relations \cref{eq:mixed_edge_vertex_commutation_relations,eq:mixed_tranfer_vertex_commutation_relations}, we introduce in \Cref{fig:graphical_notation_p1} a graphical notation where an interaction graph with nearest neighbor sites $\langle jk \rangle$ is decorated by anti-commuting fermionic operators $E_{jk}$. To account for the fact that each interaction edge $\langle jk \rangle$ corresponds to two distinct transfer operators, $T_{jk}$ and $T_{kj}$, we also define the directed graph with edges $(jk)$ and $(kj)$ decorated with the operators $T_{jk}$ and $T_{kj}$. The mixed relations of transfer operators $T_{jk}$ and $T_{kl}$ can be translated into properties of the decorated graph. Specifically, two edges connected at a node $j$ commute if there is a \emph{flow} of the corresponding two arrows (\textit{i.e.}, one arrow meets the node at its head while the other meets it at its tail), and anti-commute if there is a \emph{clash} of the arrows at this node (\textit{i.e.}, both arrows meet the node at the same end, either the head or the tail). Moreover, non-overlapping edge- and transfer- operators necessarily commute.

\section{Stabilizer-based Trotterization of fermionic Hamiltonians} \label{sec:trotterization}
In this work, we are interested in the real-time unitary evolution of two-dimensional fermionic lattice systems with maximum coordination number $n_c$, \emph{i.e.} with at most $n_c$ nearest-neighbors per site. We consider spinless Fermi-Hubbard-type Hamiltonians where the potential interaction term is diagonal in the fermionic occupation number basis. The implementation of on-site interactions is not the bottleneck of Fermi-Hubbard quantum simulations and will not be discussed in the rest of this work. We focus on the nearest-neighbor hopping contributions, given by
\begin{equation} \label{eq:tight_binding}
    H = -J \sum_{\langle jk \rangle}( c_j^\dag c_k + c_k^\dag c_j)\, ,
\end{equation}
where the sum runs over all nearest-neighbor indices $\langle jk\rangle$ on the lattice. The hopping terms are quadratic operators and can be expressed in terms of the transfer operators ($T_{jk}$),
\begin{align}
    H = J/2 \sum_{\langle jk \rangle}i(V_j - V_k)E_{jk} = J \sum_{\langle jk \rangle}(T_{jk} + T_{kj})\, .
\end{align}

\subsection{Conventional Trotterization for non-overlapping commuting operators} \label{sec:pauli_gadgets}
The hopping Hamiltonian $H$ is a free-fermionic Hamiltonian. The exact circuit implementation of the corresponding time-evolution unitary involves Givens rotations \citep{Kivlichan2018QuantumSimulationElectronicb}, or the fast-fermionic Fourier transform~\citep{Kivlichan2020ImprovedFaultTolerantQuantum} -- both operations achieving circuit depths scaling linearly with the system size. 
To achieve constant-depth implementations, one may use approximate Trotter-expansion formulas after splitting the hopping terms into smaller groups $H = H_{1}+H_{2}+\dots$, which can each be easily diagonalized separately.
These sets typically result from coloring the graph edges (see \Cref{fig:graph_coloring}\textbf{a)}) into $n_c$ color sets. 
From the mixed commutation relations of~\cref{eq:mixed_tranfer_vertex_commutation_relations}, we find that the union of non-overlapping transfer operators
\begin{equation}\label{eq:flowset_size2}
    \bigcup_{\langle jk \rangle \in \text{color}} \{ T_{jk}, T_{kj} \}\, ,
\end{equation}
forms a set of mutually commuting operators.
Hence, the corresponding time-evolution can be implemented through Trotter decompositions without incurring any Trotter error. One can then apply a (first-order) Trotter expansion on the grouping $H = \sum_{C \in \text{colors}} H_{C}$ and then exactly expand the time-evolution associated with the non-overlapping unions as
\begin{equation}
    U(dt) = \exp\left(-idt H\right) \approx \prod_{C \in \text{colors}} \exp\left(i J dt \sum_{\langle jk\rangle \in C} (T_{jk} + T_{kj}) \right)\, ,
\end{equation} 
with
\begin{equation}
    \exp\left(i Jdt \sum_{\langle jk\rangle \in C} (T_{jk} + T_{kj}) \right) = \prod_{\langle jk\rangle \in C} \exp\left(iJdt (T_{jk} + T_{kj}) \right)\, . \label{eq:color_gadgets}
\end{equation}
With this grouping strategy, it is sufficient to compile the individual gadget unitaries $U_{\langle jk\rangle} = \exp\left(iJ dt (T_{jk} + T_{kj}) \right)$ optimally to obtain efficient compilation of the global unitary evolution. This has been discussed for some common fermionic representations in \citep{Nigmatullin2025ExperimentalDemonstrationBreakeven,Jafarizadeh2025RecipeLocalSimulationa}.

\subsection{Flow-set Trotterization of overlapping commuting operators}
\begin{figure}[ht!]
    \centering
    \input{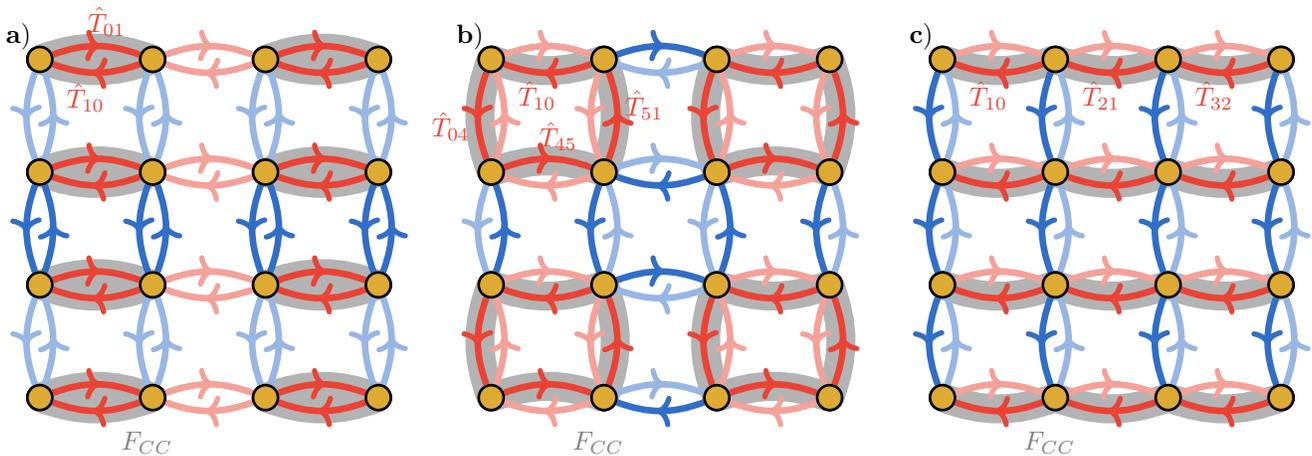}
    \caption{Three alternative grouping strategies for the tight-binding Hamiltonian terms of a $4\times 4$ square fermionic lattice. The directed edges correspond to the transfer operators $T_{jk}$ and $T_{kj}$ respectively, and the four colors (dark and light blue, dark and light red) label the four sets of commuting operators that are constructed for a square lattice. For the flow set corresponding to the dark red color, the connected components $F_{CC}$ are highlighted in gray. \textbf{a)} Grouping strategy based on unions of non-overlapping size-2 loop flow sets, e.g. $\{(01), (10)\}$. \textbf{b)} Grouping strategy based on size-$4$ plaquette flow sets, e.g. $\{(04) + (45) + (51) +(10)\}$. \textbf{c)} Third grouping strategy based on non-overlapping size-$L$ line flow sets, e.g. $\{(10) + (21) + (32) + \dots\}$, with $L$ being the lattice size.}
    \label{fig:graph_coloring}
\end{figure}

In this work, we extend the previous grouping strategy to a wider class of commuting-operator sets that we generally refer to as \emph{flow sets}. 
Flow sets generalize the construction reported above by noting that the commutation relations $[T_{jk},T_{kj}]=0$ exploited to derive \cref{eq:flowset_size2} are a special case of the mixed-commutation relations between transfer operators with chained indices $[T_{jk}, T_{kl}] = 0$ for any three indices $j,k,l$, in \cref{eq:mixed_tranfer_vertex_commutation_relations}. 
A flow set is defined by the union of non-overlapping connected components (CC), each of them being composed of transfer operators corresponding to one-dimensional subgraphs (chains or loops) of the directed interaction graph. Flow-set decompositions are valid partitions of the Hamiltonian terms into sets of commuting operators. 
For the square lattice, four flow sets are sufficient to cover the full directed graph. Example coverings for an $L\times L$ lattice are provided in \Cref{fig:graph_coloring}. In this case, natural choices are elementary size-2 flow sets, see \Cref{fig:graph_coloring}a), size-4 even and odd plaquette flow sets, see \Cref{fig:graph_coloring}b), or the size-$L$ horizontal/vertical line flow sets, see \Cref{fig:graph_coloring}c). Applying the first-order Suzuki-Trotter (ST1) formula on the flow sets then gives
\begin{equation}
\label{eq:unitary_encoding_circuit_majorana}
    \exp\left(-i dt H\right) \stackrel{ST1}{\approx} \prod_{F\in\text{flow sets}} \exp\left(iJ dt\sum_{(jk) \in F} T_{jk} \right).
\end{equation}    

Fermionic operators within a flow set $F$ commute and thus define a Majorana stabilizer group \citep{Bettaque2025StructureMajoranaClifford}. We propose to use the stabilizer formalism to execute efficiently the corresponding evolution operator. Fermionic operators belonging to different connected components (CC) of the same flow set have non-overlapping supports. Hence, the corresponding time-evolution operators can be trivially applied simultaneously. This allows us to write exactly 
\begin{equation}    
    \exp\left(iJ dt \sum_{(jk) \in F} T_{jk} \right) = \prod_{F_\text{CC} \in F} \exp\left(iJ dt \sum_{(jk) \in F_\text{CC}} T_{jk} \right) \, .
\end{equation}
Conversely, operators within the same connected component of a flow set overlap (although they still commute). It is known~\citep{Mudassar2024EncodingMajoranaCodes} that for any Majorana stabilizer group, there exists a Majorana Clifford encoding unitary $U_{F_\text{CC}}$ that encodes a product state, with eigenvalues $+1$ for all trivial stabilizers $V_l$, into a stabilizer state with $+1$ eigenvalue on all stabilizers $T_{jk}$. 
Equivalently, in the Heisenberg picture, encoding circuits map the stabilizers $T_{jk}$ in $F_{CC}$ to trivial stabilizers, the individual fermonic parities $V_l$. We denote by $l=\mathrm{enc}(jk)$ the map from the indices $(jk)$ of the stabilizers (the transfer operators) to the indices $l$ of the fermionic parity $V_l$, and stress that this map explicitly depends on the encoding circuit. When the stabilizer group is underconstrained, \emph{i.e.} the number of stabilizers is strictly smaller than the number of fermionic modes, then the stabilizer operators are mapped to a subset of fermionic modes, otherwise they are mapped one-to-one to all the fermionic parity operators. Similar strategies have been discussed in \citep{Chen2023EquivalenceFermiontoQubitMappings} to show the equivalence of fermion-to-qubit mappings up to Clifford transformation, although at the qubit level. The time-evolution unitaries for every connected components $F_\text{CC}$ of the same flow sets then can be decomposed as
\begin{equation} \label{eq:fermionic_encoding_circuit}
    \exp\left(iJ dt \sum_{(jk) \in F_\text{CC}} T_{jk} \right) = U_{F_\text{CC}}^\dagger \left( \prod_{(jk) \in F_\text{CC}} \exp(iJ dtV_{\mathrm{enc}(jk)}) \right) U_{F_\text{CC}} \, .
\end{equation}
The problem of efficiently compiling local gadget unitaries $U_{\langle jk \rangle}$ (as in \cref{sec:pauli_gadgets}) is replaced by the problem of finding flow sets that can be rotated to trivial fermionic gates $\exp(iJ dtV)$ using Majorana Clifford circuits $U_{F_{CC}}$ that can be efficiently implemented on quantum processors~\citep{Bettaque2025StructureMajoranaClifford,Mudassar2024EncodingMajoranaCodes}. 
Although the presented framework will not yield, in general, an advantage over the standard compilation approaches for \textit{any} fermion-to-qubit encoding and for \textit{any} choice of flow sets, it enables exploiting the well-established toolbox of stabilizer states and error correction in the context of Trotterized time-evolution unitaries for fermions. 
As discussed below, our approach yields especially shallow quantum circuits for the Trotter time evolution unitary whenever a two-dimensional encoding can be decomposed into flow sets with shallow encoding unitaries.

\section{Qubit-based unitary encoding of one-dimensional quadratic fermionic operators}\label{sec:qubit_representation_fermionic_operators}
For a given flow set $F$, the connected components $F_\text{CC}$ are one-dimensional subsets (lines or loops) of the system's two-dimensional directed interaction graph. For this reason, we study in \Cref{sec:representation_fermionic_quadratic_operators} fermionic quadratic operators defined on one-dimensional lattices, and in \Cref{sec:encoding_unitaries_1D} the corresponding unitary encoding circuits.
These will serve as a building blocks to construct time-evolution circuits for two-dimensional lattices.
Although so far we worked at the level of fermionic operators, we are ultimately interested in the quantum simulation of fermionic systems with simulation platforms based on qubit degrees-of-freedom~\citep{Tacchino2020QuantumComputersUniversal,Miessen2023QuantumAlgorithmsQuantuma}.
Hence, we now introduce locality-preserving fermion-to-qubit encodings~\citep{Ball2005FermionsFermionFields,Verstraete2005MappingLocalHamiltoniansa,Bochniak2020BosonizationBasedClifford,Derby2021CompactFermionQubit,Verstraete2005MappingLocalHamiltoniansa,Steudtner2019QuantumCodesQuantum,Derby2021CompactFermionQubita} to express the fermionic Hamiltonians in terms of qubit systems with $N_q$ qubits and the associated Pauli group.

\subsection{Pauli algebra and fermion-to-qubit encodings}
Many platforms for quantum simulations rely on qubits, \emph{i.e.} spin-$\frac{1}{2}$ degrees-of-freedom with hardcore-bosonic statistics. The corresponding creation and annihilation operators on site $j$ are defined as $b_j/b_j^\dag = (\sigma^x_j\pm i\sigma^y_j)/2$ in terms of the Pauli matrices ($\sigma^x_j,\sigma^y_j,\sigma^z_j$), denoted also as $(X_j,Y_j,Z_j)$ up to an irrelevant scaling factor. The Pauli $\sigma^z$ is diagonal and is generally reserved for representing the occupation number operator $n_j$. Due to its similarity to the Majorana algebra, the on-site algebra of Pauli operators
\begin{equation}
    \{\sigma^\alpha, \sigma^\beta\} = 2 \delta^{\alpha \beta} I \, ,
\end{equation}
will become a central piece of our discussion~\footnote{The algebra of the Pauli operators $\sigma^\alpha$ is not the same as the algebra of the spin creation and annihilation operators}. Any set of qubit operators, denoted with hats as $\hat{V}_k, \hat{T}_{jk}$, that are expressed as tensor products of Pauli matrices acting on qubit variables and that satisfy the relations in \Cref{eq:mixed_tranfer_vertex_commutation_relations} defines a valid fermionic encoding. 
The map $V_k, T_{jk} \mapsto \hat{V}_k, \hat{T}_{jk}$, from the fermionic operators to the qubit operators, uniquely defines an isomorphism between the even-parity observable algebra of the fermionic system and the observable algebra of the target qubit system. If the support of the qubit operators $\hat{V}_k, \hat{T}_{jk}$ does not depend on the total system size the encoding is said to be \emph{local}. 
Note that the notion of locality depends on the distance between qubits and, therefore, on the qubit lattice connectivity. 
For convenience, we restrict ourselves to mappings that encode the fermionic parity operator $V_j$ through the single-qubit Pauli operator $Z_j$ associated with the \emph{physical qubit} for the site $j$. With this choice, one finds that the Pauli string representation of edge and vertex operators $\hat{E}_{jk}$ and $\hat{T}_{jk}$ must have support at least on the physical qubits $j$ and $k$ and, on that site, the corresponding Pauli should be either $X$ or $Y$ operators in order to fulfill \cref{eq:mixed_tranfer_vertex_commutation_relations}. More general cases are explored in \cref{sec:appendix_parity_nonlocal_encodings}.

\begin{figure}[ht!]
    \centering
    \input{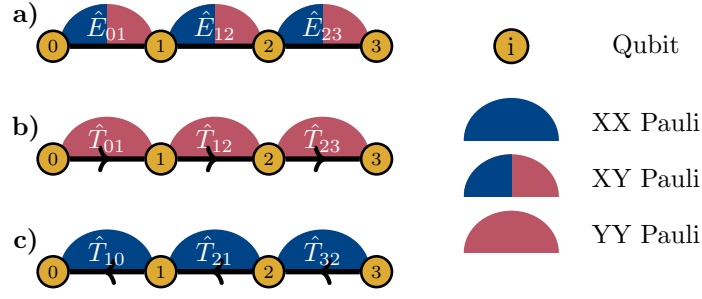}
    \caption{
    Graphical representation of Pauli string operators acting on the Hilbert space of a multi-qubit system. The three colors red, blue and purple, are used to represent the three Pauli operators $\{X,Y,Z\}$ of a qubit lattice Hamiltonian, and the corresponding surfaces cover the indices of the qubits acted upon, as in the stabilizer formalism. 
    For example the anti-commutation of the qubit operators $\hat{T}_{01}=Y_0Y_1$, in \textbf{b)}, and $\hat{T}_{21}=X_1X_2$, in \textbf{c)}, is visualized by the fact that they overlap on site 1 with different colors. The Jordan-Wigner isomorphism is represented graphically by overlaying the graphical representation of the fermionic operators (black lines) and of the Pauli operators (colored surfaces).
    }
    \label{fig:graphical_notation_p2}
\end{figure}

The Jordan-Wigner (JW) transformation is a fermionic encoding that maps a lattice of $N_f$ fermionic modes and a one-dimensional lattice of $N_q=N_f$ qubits. 
For fermionic Hamiltonians defined on a one-dimensional lattice, JW leverages the local qubit Pauli algebra at site $j$, whose states store the $j$-th fermionic parity, to impose the anti-commutation of two consecutive edge operators $E_{j-1,j}$ and $E_{j,j+1}$ as in ~\cref{eq:mixed_tranfer_vertex_commutation_relations}. 
The operator isomorphism associated with the JW mapping is defined by the following qubit operators,
\begin{align}
    \hat{V}_j &= Z_j\\
    \hat{E}_{j,j+1} &= X_j Y_{j+1} \\
    \hat{T}_{j,j+1} &= -\tfrac{1}{2} Y_j Y_{j+1}\, .\label{eq:jw_isomorphim}
\end{align} 
For readability, in \Cref{fig:graphical_notation_p2} we introduce a graphical notation for tensor products of Pauli matrices. This new notation is complementary to the one of \Cref{fig:graphical_notation_p1} for fermionic quadratic operators. The three Pauli operators $X, Y, Z$ are represented by the three colors: blue, red and purple, respectively and we do not visually account for the prefactors in \cref{eq:jw_isomorphim}. We further use the convention of the Clifford stabilizer formalism, where a Pauli string operator acting on multiple sites $j,k,l,\dots$ is represented by a multi-colored surface with overlap on the sites $j,k,l,\dots$. The context in which we use the stabilizer formalism is however not to be confused with the stabilizer/gauge constraints that naturally appear in two-dimensional local encodings. The anti-commutation of two Pauli-strings is then visually determined by the number of color \emph{clashes} in the overlap of the two colored surfaces. Any overlay of fermionic and qubit operator representations implies a fermion-to-qubit encoding, \emph{e.g.} an overlay of a fermionic edge $E_{jk}$ colored in black and a qubit operator $\hat{E}_{jk}$ acting on the corresponding qubit vertices $j,k$ implies a qubit representation of the fermionic edge as $E_{jk}\mapsto \hat{E}_{jk}$.

\subsection{Representation of fermionic quadratic operators on 1D lattices}\label{sec:representation_fermionic_quadratic_operators}

Only three anticommuting qubit operators can be defined from the on-site Pauli algebra of a single qubit. As the individual fermionic parities are encoded as $\hat{V}_k=Z_k$, only two fermionic anticommuting edges intersecting at $k$ can be represented with Pauli operators.
For square fermionic lattices, one needs to provide the qubit representation of four fermionic edge operators intersecting at a fermionic site $k$.
Therefore, it is not possible to leverage the \textit{local} Pauli algebra at site $k$ to enforce all anticommutation relations.
Local fermionic mappings in two dimensions address this problem by introducing ancilla qubits in addition to the $k$-th physical qubit. In practice, when restricted to one-dimensional sublattices, two-dimensional encodings induce \emph{quasi} one-dimensional encodings that differ from JW by the inclusion of these ancilla degrees-of-freedom.
We now propose a categorization of all such representations of one-dimensional fermionic lattices defined on top of the JW encoding.

\begin{figure*}[ht!]
    \centering
    \input{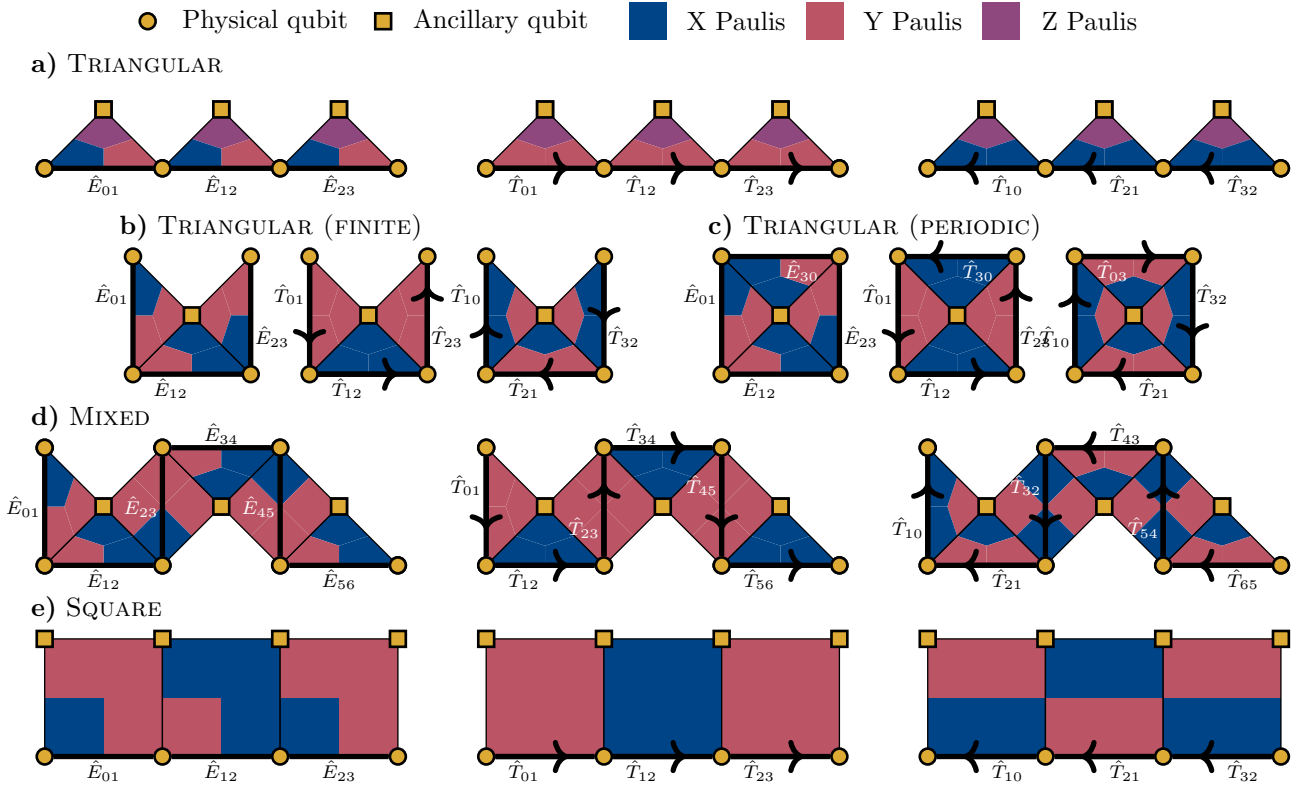}
    \begin{tikzpicture}[
    x=\papertwocolumnwidth/46,
    y=\papertwocolumnwidth/46,
    every edge/.append style={
        color=black,
        line width = \halfedgelinewidth, 
    },    
    every path/.append style={
        color=black,
        line width = 0.5, 
    },
    every node/.style={
        text=white, 
        scale = 1,
        line width = \smallnodelinewidth, 
        inner sep=0,
        minimum size = \smallnodeminumumsize, 
        font=\nodefontsize
        },
        decoration={markings, mark= at position 0.60 with {\arrow[scale=1]{>}},}
        ]

        \coordinate (LEGEND) at (3.5, 4.5);
        \draw (LEGEND) pic {legend};
        
        
        \coordinate (PLOTE) at (0, 3);
        \node[above, xshift=0, yshift=0,color=black, align=left, anchor=west, font=\legendfontsize] at ($(PLOTE) + (-0.5 , 0)$) {\textbf{a)} \textsc{Triangular}};
        \draw ($(PLOTE) + (0 , -3.5)$) pic {jordanwigner1d_edges_idleancillas_everyone_v3};
        \draw ($(PLOTE) + (15, -3.5)$) pic {jordanwigner1d_transfersTij_idleancillas_everyone};
        \draw ($(PLOTE) + (30, -3.5)$) pic {jordanwigner1d_transfersTji_idleancillas_everyone};
        
        \coordinate (PLOTA) at (3, -4);
        \node[above, xshift=0, yshift=0,color=black, align=left, anchor=west, font=\legendfontsize] at ($(PLOTA) + (-0.5,1.5)$) {\textbf{b)} \textsc{Triangular (finite)}};
        \draw ($(PLOTA) + (0, -3.5)$) pic {jordanwigner1d_edges_1ancilla};
        \draw ($(PLOTA) + (6, -3.5)$) pic {jordanwigner1d_transfersTij_1ancilla};
        \draw ($(PLOTA) + (12,-3.5)$) pic {jordanwigner1d_transfersTji_1ancilla};
        
        \coordinate (PLOTB) at (23, -4);
        \node[above, xshift=0, yshift=0,color=black, align=left, anchor=west, font=\legendfontsize] at ($(PLOTB) + (-0.5,1.5)$) {\textbf{c)} \textsc{Triangular (periodic)}};
        \draw ($(PLOTB) + (0 , -3.5)$) pic {jordanwigner1dperiodic_edges_1ancilla};
        \draw ($(PLOTB) + (6 , -3.5)$) pic {jordanwigner1dperiodic_transfersTij_1ancilla};
        \draw ($(PLOTB) + (12, -3.5)$) pic {jordanwigner1dperiodic_transfersTji_1ancilla};

        
        \coordinate (PLOTC) at (0, -10.5);
        \node[above, xshift=0, yshift=0,color=black, align=left, anchor=west, font=\legendfontsize] at ($(PLOTC) + (-0.5, 1.5)$) {\textbf{d)} \textsc{Mixed}};
        \draw ($(PLOTC) + (0 , -3.5)$) pic {jordanwigner1d_edges_ancillas_everytwo_v3};
        \draw ($(PLOTC) + (15, -3.5)$) pic {jordanwigner1d_transfersTij_ancillas_everytwo};
        \draw ($(PLOTC) + (30, -3.5)$) pic {jordanwigner1d_transfersTji_ancillas_everytwo};

        \coordinate (PLOTD) at (0, -17);
        \node[above, xshift=0, yshift=0,color=black, align=left, anchor=west, font=\legendfontsize] at ($(PLOTD) + (-0.5, 1.5)$) {\textbf{e)} \textsc{Square}};
        \draw ($(PLOTD) + (0 , -3.5)$) pic {jordanwigner1d_edges_ancillas_everyone_v3};
        \draw ($(PLOTD) + (15, -3.5)$) pic {jordanwigner1d_transfersTij_ancillas_everyone};
        \draw ($(PLOTD) + (30, -3.5)$) pic {jordanwigner1d_transfersTji_ancillas_everyone};
        
\end{tikzpicture}
    \caption{Edge and transfer operator representation for a one-dimensional chain of fermions beyond Jordan-Wigner. \textbf{a)}~Example construction with weight-3 \emph{triangle-shaped} transfer operators. Even though this construction introduces ancilla degrees of freedom, the anti-commutation relations are encoded through the physical qubits as in JW. \textbf{b)} and \textbf{c)} One ancilla qubit is sufficient to enforce the mixed relations of three consecutive edges, and alternatively of four consecutive edges with periodic boundary conditions. In both cases, the transfer operators are represented by weight-3 \emph{triangular-shaped} Pauli operators. In~\textbf{d)} an ancilla qubit is added every second physical qubit while in \textbf{e)} an ancilla is added for every physical qubit. The transfer operators are \emph{mixed-shaped} weight-3 or 4 operators in \textbf{d)} and weight-4 \emph{square-shaped} operators in \textbf{e)}.}
    \label{fig:flavors_of_jw}
\end{figure*}

A first nontrivial extension of the JW mapping consists of including the ancilla qubits in the representation of the transfer operators, while still enforcing the edge anti-commutation relations (ACRs) through Pauli operators defined on physical qubits:
\begin{itemize}
    \item \textbf{Physical qubits enforcing all ACRs.} An example is represented in \Cref{fig:flavors_of_jw}\textbf{a)}, where an ancilla qubit is included for each physical edge. This ancilla qubit does not enforce any anti-commutation relations. Despite the similarity with the plain one-dimensional Jordan-Wigner encoding, we will see that the increased Pauli weight allows for more efficient circuit realization of the time-evolution operator.
\end{itemize}
We already mentioned that, for two-dimensional encodings, it is impossible to enforce all ACRs only using the physical qubits. Ancillary degrees-of-freedom allow for new ways to enforce the ACRs. Specifically:
\begin{itemize}
    \item \textbf{A single ancilla can enforce ACRs for up to 3 (or 4) consecutive edges.} 
    To do so, the support of each edge operator is increased to include the corresponding ancillary degree of freedom. Instead of anti-commuting because of the Pauli operators defined on the physical qubits, two consecutive edges are made to anti-commute through the Pauli operator defined on the ancilla qubit, see \Cref{fig:flavors_of_jw}\textbf{b)}. The same applies for a line of four edges with periodic boundary conditions, see \Cref{fig:flavors_of_jw}\textbf{c)}. In both cases, the edge and transfer operators are mapped onto weight-3 Pauli operators.
\end{itemize}
To extend this construction to infinite one-dimensional lattices, one must regularly introduce ancillary degrees-of-freedom, as each ancilla can enforce the ACRs of either two or three consecutive edges:
\begin{itemize}
    \item \textbf{Ancillas enforcing ACRs of 3 consecutive edges.} In \Cref{fig:flavors_of_jw}\textbf{d)}, ancillary qubits are added every second edge to locally enforce the mixed relations of \cref{eq:mixed_tranfer_vertex_commutation_relations} for three consecutive edges. At the boundary of each of these local patches, we find that the weight of edge and transfer operators must be increased to 4 in order to satisfy the mixed relations with both the prior and later edges. 
    \item \textbf{Ancillas enforcing ACRs of 2 consecutive edges.} While one ancillary qubit can be used to enforce the pairwise anti-commutation of up to three edges on an infinite lattice, it may be preferable for some applications to introduce more ancillary qubits, each enforcing a smaller number of ACRs. One can use a single ancillary qubits per physical qubit (except from the boundaries), each accounting for the local mixed relations of two consecutive edges only. In this case, depicted in~\Cref{fig:flavors_of_jw}\textbf{e)} the weight of the edge- and transfer- operators in all cases increases to 4 (except at the boundaries). 
\end{itemize}

\subsection{Clifford encoding unitaries for 1D stabilizer states} \label{sec:encoding_unitaries_1D}
\begin{figure*}[ht!]
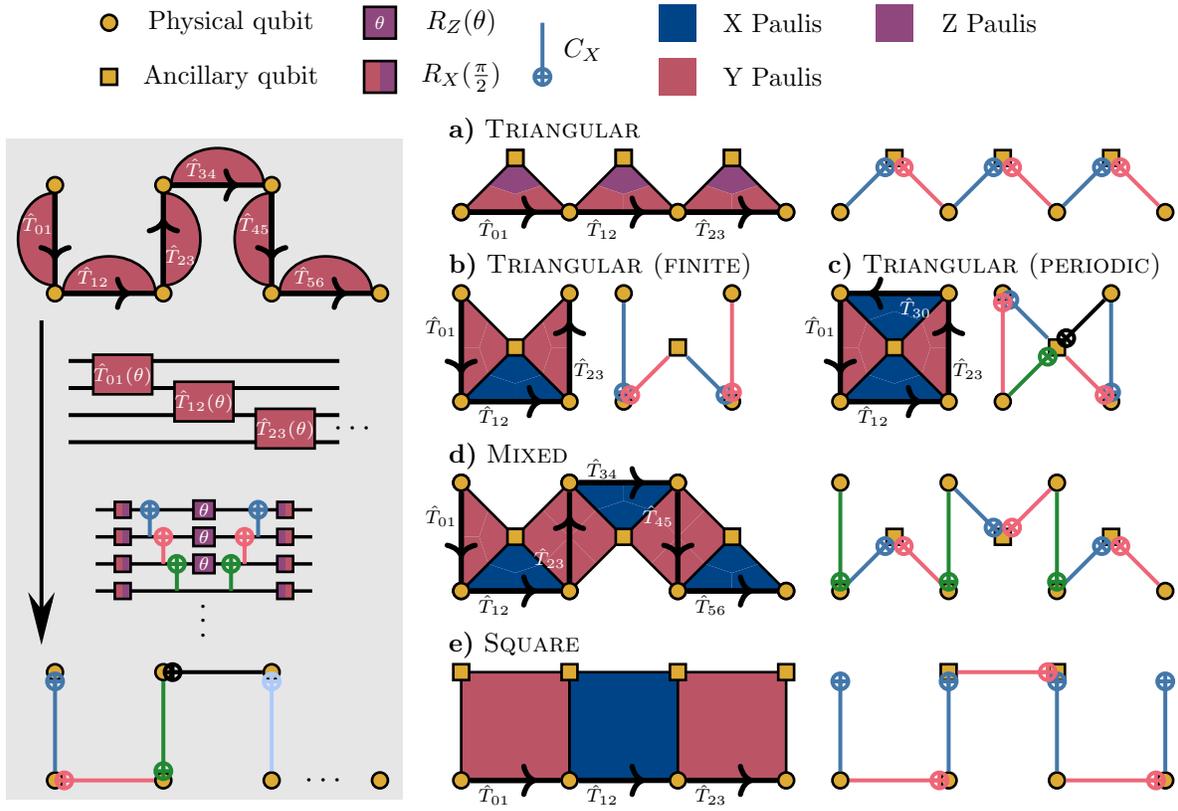

    \centering
    \input{tikz_final/FIG04_Representation_Quadratic_Operators/FIG04_Representation_Quadratic_Operators_utils.tikz}
    \input{tikz_final/FIG05_Clifford_Unitaries/FIG05_Clifford_Unitaries.tikz}
    \caption{Clifford encoding circuits for all of the examples enumerated in \cref{sec:representation_fermionic_quadratic_operators}. The Clifford encoding unitaries are given by sequences of entangling $CX$ gates layered according to the colors: blue, red, green, black, cyan. \emph{Left}. Derivation of the parametrized time-evolution unitaries for the Jordan-Wigner encoding. For this example only we explicitly include the layer of single-qubit basis rotations that needs to be applied before the encoding circuit. For the Jordan-Wigner case, the encoding depth scales linearly with the system size and each layer (color) only contains a single entangling gate. \emph{Right}. We show in \textbf{a)} that increasing the weight the Jordan-Wigner transfer-operators by introducing ancilla qubits yields a constant-depth preparation circuit, in contrast to the linear-depth baseline circuit. \textbf{b)}~Depth-2 encoding circuit for three edges with one ancilla register. The set of stabilizers in \textbf{c)} corresponds to that of the surface code for $L=2$~\citep{Higgott2021OptimalLocalUnitary} and is implemented in depth-4. For infinite one-dimensional lattices, we find encoding unitaries of depth-3 for our mapping with qubit-to-fermion ratios 3-to-2 (\textbf{d)}) and depth-2 for qubit-to-fermion ratios 2-to-1 (\textbf{e)}).}
    \label{fig:clifford_circuit}
\end{figure*}

After having described the qubit representations of fermionic transfer operators for one-dimensional lattices, we now derive the qubit representations of the associated fermionic Clifford encoding circuits $U_{F_{CC}}$ of \cref{eq:fermionic_encoding_circuit}. We denote with $\hat{U}_{F_\text{CC}}$ the qubit unitary obtained by applying the fermion-to-qubit transformation to the fermionic unitary encoding. In the language of the Pauli stabilizer formalism~\citep{Gottesman1997StabilizerCodesQuantum}, $\hat{U}_{F_\text{CC}}$ is the Clifford encoding circuit associated with the qubit stabilizer group $\{\hat{T}_{jk}, (jk) \in F_\text{CC}\}$. For each qubit representation of fermionic quadratic operators on one-dimensional lattices categorized in \cref{sec:representation_fermionic_quadratic_operators}, we present in \Cref{fig:clifford_circuit} a shallow-depth Clifford encoding circuit. Here, the shallowness is measured by the number of entangling $CX$ layers and represents the degree to which quantum operations can be applied in parallel on a digital quantum processor. 

As a reference, we consider the conventional JW mapping. In this case, the commuting transfer operators are mapped to single-qubit operators via a ladder of $CX$ gates whose depth scales linearly with the system size (see \Cref{fig:clifford_circuit}, left). From the first encoding discussed in \Cref{sec:representation_fermionic_quadratic_operators}, we highlight how adding ancillary degrees of freedom and increasing the weight of Pauli operators improves the depth of their quantum circuit realization. In \Cref{fig:clifford_circuit}\textbf{a)}, we show how augmenting a Jordan-Wigner-type encoding by adding one ancillary degree of freedom for each edge reduces the depth of the optimal Clifford encoding circuit to 2, compared to the linearly-scaling depth of the initial construction. We define the notion of depth-optimality for a stabilizer encoding circuit based on the Pauli weight of the stabilizers. For instance, weight-2 Paulis require a Clifford preparation circuit of at least depth-1, while both weight-3 and weight-4 Paulis require a Clifford preparation circuit of at least depth-2~\footnote{Note that for simplicity, the provided circuits prepare the stabilizers in the $X,Z$ basis instead of the $X,Y$ basis. The two are connected by a layer of single qubit rotations.}.

For all the other representations of one-dimensional fermionic quadratic operators defined in \Cref{sec:representation_fermionic_quadratic_operators}, we find that the depth is mainly determined by the number of anti-commutation relations accounted for by ancillary degrees-of-freedom, which induce sequential circuit representations instead of dense and shallow representations. Noteworthy, the configuration  given in \Cref{fig:clifford_circuit}\textbf{c)}, with one ancilla qubit enforcing the anti-commutation of four edges on a periodic lattice, corresponds exactly to a toric-code state preparation circuit with depth-4 \citep{Higgott2021OptimalLocalUnitary}. For the representation with a qubit-to-fermion ratio of 3-to-2, we propose in \Cref{fig:clifford_circuit}\textbf{d)} a depth-3 encoding circuit, which reduces to a depth-2 circuit when only three edges are considered, see \Cref{fig:clifford_circuit}\textbf{b)}. Notably, we derive an optimal depth-2 encoding circuits for the fermionic encoding with the largest qubit-to-fermion ratios of 2-to-1, see \Cref{fig:clifford_circuit}\textbf{e)}. These results highlight that a trade-off between \emph{space} and \emph{time} complexities exists: larger qubit-to-fermion ratios lead to more dense and shallow time-evolution quantum circuits, since operations can be applied in parallel.

\section{Stabilizer-based Trotterization for 2D fermionic simulations}\label{sec:clifford_pauli_encoding}
The Trotterization strategy introduced in \cref{sec:trotterization} for two-dimensional lattice Hamiltonians can be applied to any chosen combination of fermion-to-qubit mappings and flow sets. Standard approaches for constructing quantum circuit representation of the Trotter fermionic dynamics use the size-2 petal flow sets of \cref{fig:graph_coloring}\textbf{a)} and optimize the corresponding Pauli unitary gadgets $\hat{U}_{\langle jk\rangle}=\exp(iJdt (\hat{T}_{jk}+\hat{T}_{jk}))$ individually. We argue in this section that larger flow sets, such as the plaquette or line flow sets of \cref{fig:graph_coloring}\textbf{b)c)} are better suited than common two-dimensional fermion-to-qubit mappings proposed in the literature as they yield shallower circuit representations. As a test-case, we use the widely studied Verstraete-Cirac (VC) mapping~\citep{Verstraete2005MappingLocalHamiltoniansa} of a square fermionic lattice. Additional details on this mapping, as well as on how other common mappings of the literature can be included in our framework, are reported in \cref{sec:appendix_paritylocal_encoding}. 

\subsection{Flow set decomposition for the VC mapping}\label{sec:choosing_flowsets}
The VC mapping~\citep{Verstraete2005MappingLocalHamiltoniansa} is a parity-local mapping with a qubit-to-fermion ratio of 2-to-1, and weight-3 and weight-4 transfer operators. The properties of this mapping are detailed in \cref{sec:appendix_vc_encoding}, and the relevant transfer operator are depicted in \Cref{fig:flowset_compilation_vc_p1}. An efficient flow set decomposition of hopping operators in the VC mapping satisfies two properties: first it uses the most shallow encoding unitaries from \cref{sec:encoding_unitaries_1D}, and second it guarantees that the connected components of each flow set remain non-overlapping once the fermion-to-qubit mapping is applied. The latter condition guarantees the maximum degree of parallelization of the corresponding quantum circuit. Both properties are satisfied for the petal flow sets as well as for the line flow sets of \cref{fig:graph_coloring}, which both group the horizontal hoppings into triangular-type CCs and vertical hoppings into square-type CCs.
With the second choice, the Suzuki-Trotter fermionic evolution \cref{eq:unitary_encoding_circuit_majorana} over the four line flow sets oriented to the east (EA), the west (WE), the north (NO), and the south (SO), becomes 
\begin{align}\label{eq:vc_trotter_formula}
    \exp(-idt H) &\stackrel{ST1}{\approx} \exp(-idt H_{NO})\exp(-idt H_{SO})\exp(-idt H_{WE})\exp(-idt H_{EA})\, .
\end{align}

Moreover, for fermion-to-qubit encodings with on-site parity operators $\hat{V}_j=Z_j$, the two orientations for both vertical and horizontal transfer operators are related through a layer of single-qubit rotations. For the specific case of the VC encoding, the fermionic product equality $T_{kj} = -V_j V_k T_{jk}$ then becomes at the qubit level,
\begin{equation}
    \hat{T}_{jk} = -Z_j Z_k \hat{T}_{kj} = [R_j^Z(\tfrac{\pi}{2}) R_k^Z(\tfrac{\pi}{2})] \,\hat{T}_{kj} \,[R_j^Z(\tfrac{\pi}{2}) R_k^Z(\tfrac{\pi}{2})]^\dag\, ,\label{eq:tjk_tkj_identity}
\end{equation}
where, in the last equality, we used the fact that for the VC mapping (see \cref{eq:VC_transfer_operators}) the support of $\hat{T}_{kj}$ on the physical qubits is $X_k X_j$ or $Y_k Y_j$, and that $Z_l X_l = -i R_l^Z(\tfrac{\pi}{2}) X_l R_l^Z(\tfrac{\pi}{2})^\dag$. After the fermion-to-qubit mapping, the pairs of flow sets $(\hat{H}_{NO}, \hat{H}_{SO})$ and $(\hat{H}_{WE}, \hat{H}_{EA})$ are then related by a layer of single-qubit~$R_l^Z(\pi)$ gates acting on the physical qubits
\begin{align}\label{eq:trotter_hdown}
    \exp(-idt \hat{H}_{SO}) = \left(\prod_{l\in \text{phys.}} R_l^Z(\tfrac{\pi}{2}) \right) \exp(-idt \hat{H}_{NO}) \left(\prod_{l\in \text{phys.}} R_l^Z(\tfrac{\pi}{2}) \right)^\dag\, , \\
    \exp(-idt \hat{H}_{WE}) = \left(\prod_{l\in \text{phys.}} R_l^Z(\tfrac{\pi}{2}) \right) \exp(-idt \hat{H}_{EA}) \left(\prod_{l\in \text{phys.}} R_l^Z(\tfrac{\pi}{2}) \right)^\dag\, . \label{eq:trotter_heast}
\end{align}
Hence, decomposing the Verstraete-Cirac encoding along the line flow sets of \cref{fig:graph_coloring} only requires circuit representations of two independent Clifford encoding circuits. In a Trotter evolution circuit such as \cref{eq:vc_trotter_formula}, these are parallelized across all connected components and conjugated with layers of single qubit Paulis to implement the hoppings on all horizontal and vertical lines.

\subsection{Time-evolution unitaries for the VC mapping} \label{sec:composing_1D_encoding_circuits_to_2D}
We conclude our discussion by characterizing the two Clifford encoding circuits that are required to implement $\exp(-idt \hat{H}_{EA})$ and $\exp(-idt \hat{H}_{NO})$. For this, we use the classification of \cref{sec:representation_fermionic_quadratic_operators} and show in \cref{sec:appendix_vc_encoding} that the vertical (resp. horizontal) flow sets, up to single-qubit rotations, are based on the square type (resp. triangle-type) one-dimensional stabilizer group representations. 
\begin{figure}[ht!]
    \centering
    \includegraphics[width=0.8\linewidth,page=1]{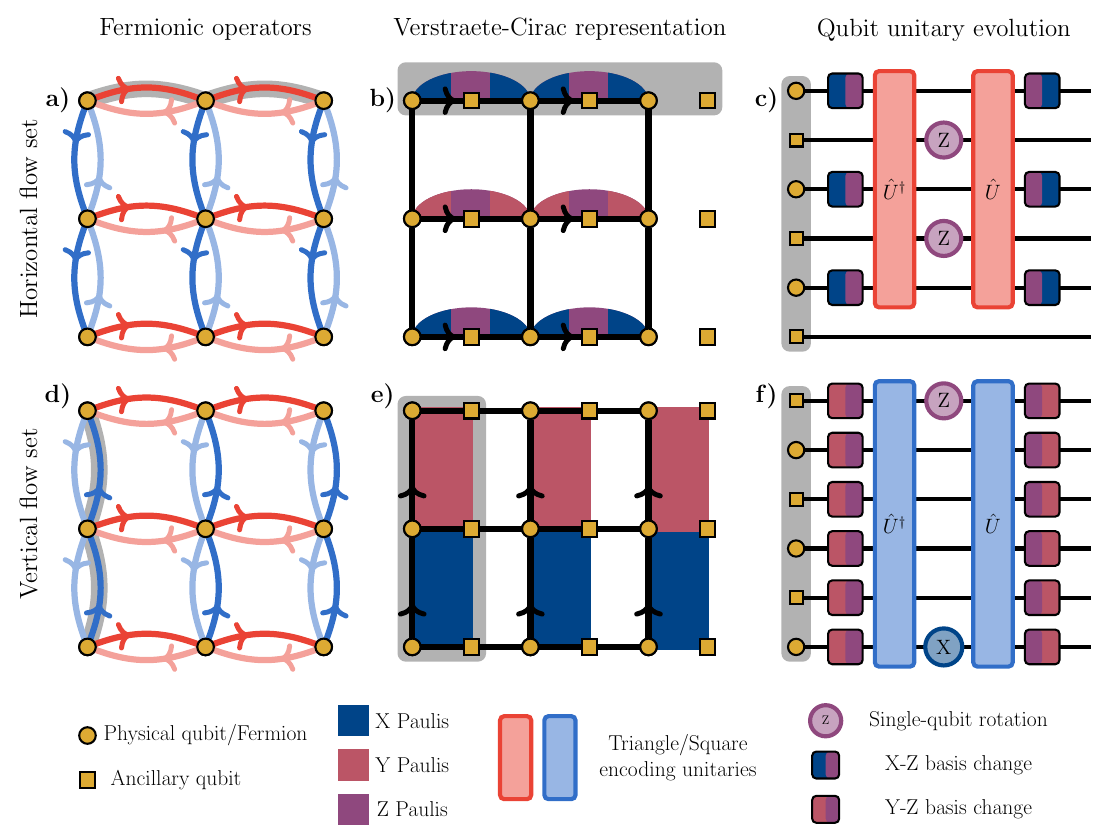}
    \caption{Stabilizer-based compilation of the fermionic time-evolution unitary under the Verstraete-Cirac mapping. \textbf{a)d)}~Horizontal and vertical line flow set decomposition of transfer operators. \textbf{b)e)}~The qubit representations of horizontal transfer operators are weight-3 Paulis that correspond to the triangular-shaped generic one-dimensional encoding found in \cref{sec:representation_fermionic_quadratic_operators}. For the vertical operators, we obtain weight-4 Paulis that correspond to the square-shaped generic encoding. \textbf{c)f)}~Triangular-type and Square-type Clifford encoding circuit applied to each connected components of the horizontal and vertical flow sets.}
    \label{fig:flowset_compilation_vc_p1}
\end{figure}
These findings are summarized in \Cref{fig:flowset_compilation_vc_p1}, where the time-evolution of non-overlapping flow sets in the VC encoding is expressed using the unitaries $\hat{U}_\text{square}$ and $\hat{U}_\text{triangle}$ introduced in \Cref{fig:clifford_circuit} and using single qubit rotations $R_l^Z(\theta)$
\begin{align}\label{eq:vc_encoding_tij_tji_identity}
    \exp(-idt \hat{H}_{NO}) & = \hat{U}_\text{square} \left(\prod_{(jk) \in F_\text{CC}} R^Z_{\mathrm{enc}(jk)}(-Jdt) \right)\hat{U}_\text{square}^\dag\, ,\\
    \exp(-idt \hat{H}_{EA}) & = \hat{U}_\text{triangle} \left(\prod_{(j'k') \in F'_\text{CC}} R^Z_{\mathrm{enc'}(j'k')}(-Jdt) \right) \hat{U}_\text{triangle}^\dag\, .
\end{align}
The maps $(jk)\mapsto \mathrm{enc}(jk)$ and $(j'k')\mapsto \mathrm{enc'}(j'k')$ defining the indices of the single-qubit rotations can be inferred from the square (resp. triangle) encoding circuits $\hat{U}_\text{square}$ (resp. $\hat{U}_\text{triangle}$) as detailed in \cref{sec:appendix_vc_encoding}. As an abuse of notation, we use the same symbols $\hat{U}_\text{square}$ and $\hat{U}_\text{triangle}$ for the same unitaries repeated across the different connected components of the flow sets. Quite notably, the Trotter unitary for the VC encoding can then be expressed as a composition of four flow sets for which we found two independent depth-optimal encoding circuits. 

At this stage, the Trotter depth is easy to calculate as each of the four flow sets can be applied with one encoding and one decoding circuit, both given in~\Cref{fig:clifford_circuit} as depth-2 circuits. Overall, our proposed Trotterization scheme then yields a depth-16 quantum circuit representation for a single Trotter step in the VC encoding. See~\cref{sec:appendix_vc_encoding} for a comparison with other approaches. This value for the depth is a higher-bound estimate that could be further reduced by circuit-level optimization at the interface of each encoding and decoding circuits; nevertheless, it is already below what existing compilation schemes achieve~\citep{Algaba2024LowdepthSimulationsFermionicb,Cade2020StrategiesSolvingFermiHubbarda,Nigmatullin2025ExperimentalDemonstrationBreakeven}. The VC mapping example shows how choosing larger flow sets and using the stabilizer formalism can yield low-depth qubit circuit decompositions of the fermionic unitary evolution. Our comparison against standard compilation schemes is detailed in \Cref{sec:appendix_vc_encoding} and holds under the assumption of a digital quantum simulator platform with a square connectivity graph and with the $CX$ being the only native entangling gate. Relaxing these constraints, as is done in \cite{Cade2020StrategiesSolvingFermiHubbarda,Algaba2024LowdepthSimulationsFermionicb} would lead to further simplifications which are not the subject of this work.

\section{Conclusion}

In this work, we propose a novel framework for quantum simulations of fermionic systems leveraging stabilizer-state encoding circuits.
Our approach is constructed around two key observations.

First, in a fermionic lattice Hamiltonian, commuting hopping terms are either non-overlapping, or can be grouped into \emph{flow sets}, which are one-dimensional subgraphs of the directed interaction graph. This observation can be leveraged when constructing the corresponding time-evolution operator: (i) Trotter-Suzuki factorizations are used to separate the evolutions of the terms in different one-dimensional subgraphs, while (ii) the evolution under commuting operators within a single subgraph, which form a stabilizer group, relies on fermionic stabilizer-encoding circuits.
    
Second, multi-dimensional local fermion-to-qubit mappings, when restricted to one-dimensional subgraphs, take simple representations which we categorize. For each of these representations, we construct a low-depth qubit stabilizer-encoding circuit. These circuits can be composed to construct efficient qubit unitary circuit representations of the fermionic evolution unitaries. Fermion-to-qubit mappings that can be decomposed using the simplest representations among the ones we categorized are appealing candidates for fermionic real-time evolution on current quantum hardware. Notably, this observation does not depend on the Pauli-weight of the encoded operators.

We show how our approach can be applied to the most common fermion-to-qubit encodings of the literature, and generally yields shallow and structured Trotter circuit decompositions. Our framework includes also the more standard Trotterization schemes, which correspond to commuting groups defined from size-2 flow sets. However, circuit realizations of these schemes must be optimized on a case-by-case level, and these optimizations become less efficient for large Pauli-weight representations of the hopping (transfer) operators. Conversely, we find that, in the Verstraete-Cirac mapping, decomposing the hopping terms in larger flow sets, \emph{i.e.} larger commuting groups, yields very efficient circuit implementations with little to no circuit-level optimization. Remarkably, a reduction in the Pauli weight of the mapped transfer operators does not, by itself, guarantee the lowest achievable Trotter depth. 
Instead, the Trotter depth in our scheme is directly related to the one-dimensional stabilizer groups underlying the fermion-to-qubit encoding. Our characterization of one-dimensional stabilizer groups then reveals a trade-off between \emph{space}, the number of auxiliary qubits required in the encoding, and \emph{time}, the parallelizability of encoded Trotter steps.

In future works, we plan to experimentally realize the real-time evolution of Fermi-Hubbard-like Hamiltonians on digital quantum platforms~\citep{Alam2025ProgrammableDigitalQuantum,Cade2020StrategiesSolvingFermiHubbarda,Alam2025FermionicDynamicsTrappedion,Granet2025SuperconductingPairingCorrelations}. Given that the hopping terms by themselves produce a solvable free-fermion evolution, we plan to apply our strategy instead to Hamiltonians incorporating non-zero on-site interactions. 
In this regime, our approach can be seen as an intermediate between the Fourier transformation circuits that globally diagonalize the kinetic terms, and the local compilation approach where small groups of hopping operators are independently diagonalized. 
Because shallow time evolution unitaries, rather than low-weight encoded operators, ultimately determine the practicality of a fermion-to-qubit mapping, we argue that the Verstraete-Cirac~\citep{Verstraete2005MappingLocalHamiltoniansa} or the Generalized Superfast encodings~\citep{Setia2019SuperfastEncodingsFermionic} might be better suited to current quantum processors than the compact Derby-Klassen~\citep{Derby2021CompactFermionQubit} or supercompact~\citep{Chen2023EquivalenceFermiontoQubitMappings} mappings that have been used in recent large-scale experiments.

\begin{acknowledgments}
We thank Sergey Bravyi for interesting discussions and suggestions. We thank Vojt{\v{e}}ch  Havl{\'\i}{\v{c}}ek and Bryce Fuller for relevant comments on the manuscript. This research was supported by the Swiss National Center of Competence in Research (NCCR) MARVEL and NCCR SPIN, funded by the Swiss National Science Foundation under grant numbers 205602 and 180604, as well as by the RESQUE project, also funded by the Swiss National Science Foundation (grant number 225229).
\end{acknowledgments}

\bibliography{biblio}

\clearpage


\appendix
\section{Application to parity-local 2D encodings}
\label{sec:appendix_paritylocal_encoding}
In this section, we apply the flow-set framework to common fermion-to-qubit mappings of the literature where the parity operator is $1$-local, \textit{i.e.}, that encode the fermionic parity $V_j$ as the single-qubit operator $Z_j$. Moreover, we compare this framework with state-of-the-art approaches that have pursued optimization of the unitary time-evolution circuits for these mappings.

\subsection{Detailed derivation for the Verstraete-Cirac encoding}
\label{sec:appendix_vc_encoding}
The Verstraete-Cirac encoding~\cite{Verstraete2005MappingLocalHamiltoniansa} is a fermion-to-qubit transformation with qubit-to-fermion ratio 2-to-1. It associates each fermionic site $j$ to a physical qubit $j$ that encodes the parity, and to an auxiliary qubit $a(j)$ that enforces anti-commutation relations. The corresponding isomorphism is given, up to single-qubit rotations, by the map
\begin{align}
    \hat{V}_{j} &= Z_{j}\\
    \hat{T}_{jk} &= \begin{cases}
        (X_{j} X_{k}) Z_{a(j)} & \text{if $(jk)$ is an horizontal edge and $j$ is on an even row}\, ,\\
        (Y_{j} Y_{k}) Z_{a(j)} & \text{if $(jk)$ is an horizontal edge and $j$ is on an odd row}\, ,\\
        (X_{j} X_{k}) X_{a(j)} X_{a(k)} &  \text{if $(jk)$ is a vertical edge and $j$ is on an even row}\, ,\\
        (Y_{j} Y_{k}) Y_{a(j)} Y_{a(k)} & \text{if $(jk)$ is a vertical edge and $j$ is on an odd row}\, .
    \end{cases} \label{eq:VC_transfer_operators}
\end{align}

\begin{figure}[ht!]
    \centering
    \includegraphics[width=\linewidth,page=2]{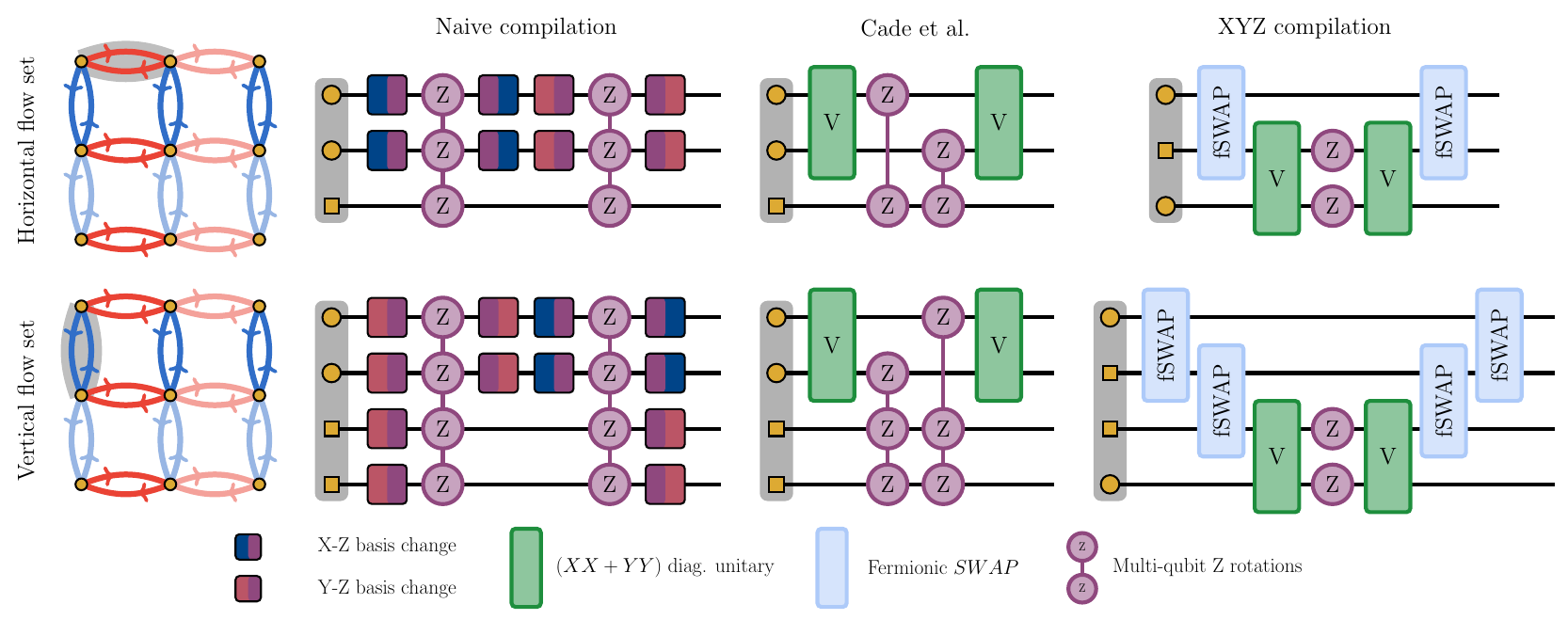}
    \caption{Standard compilation strategies can be seen as optimization of the Pauli unitary gadgets $U_{\langle jk\rangle} = \exp(-i\theta (\hat{T}_{jk}+\hat{T}_{kj}))$ for the the size-2 petal flow sets $\{T_{jk}, T_{kj}\}$. The naive compilation circuit requires two weight-3 $R_{ZZZ}$ rotations for the vertical gadgets, and two weight-4 $R_{ZZZZ}$ rotations for the horizontal gadgets. This corresponds to depth-8 and depth depth-12 circuit respectively. Alternatively, the approach of Cade et al. \citep{Cade2020StrategiesSolvingFermiHubbarda} lowers both $CX$-depths to 6 but assumes all-to-all qubit connectivity. Finally, the $XYZ$ compilation scheme of \citep{Algaba2024LowdepthSimulationsFermionicb} achieves depth-8 and depth-14 circuits for the horizontal and vertical hoppings. When assuming native $fSWAP$-gates, the $XYZ$ further improves the compiled circuit depths.}
    \label{fig:flowset_compilation_vc_p2}
\end{figure}

Conventionally, the Trotterization of the qubit-Hamiltonian resulting from the Verstraete-Cirac fermion-to-qubit encoding uses the size-2 petal flow sets of \cref{fig:graph_coloring}.
The corresponding Pauli unitary gadgets $U_{\langle jk\rangle}=\exp(iJdt (\hat{T}_{jk}+\hat{T}_{jk}))$ are then optimized individually. For example, in a naive compilation strategy, depicted in~\cref{fig:flowset_compilation_vc_p2}, the two contributions $\hat{T}_{jk}$ and $\hat{T}_{kj}$ are split by a Suzuki-Trotter factorization, which yields a depth-8 (resp. depth-12) circuit for the even and odd horizontal (resp. vertical) hoppings. Alternatively, the authors in \citep{Cade2020StrategiesSolvingFermiHubbarda} proposed to compile the evolution unitary by first employing a diagonalization circuit to simplify the $(XX+YY)$ Pauli appearing in the $(\hat{T}_{jk}+\hat{T}_{jk})$ term, followed by diagonal two-qubit $R_{zz}$ rotations. However, this approach requires an increased qubit connectivity, and is therefore not suitable for hardware platforms with square qubit topology. At last, the recently-proposed $XYZ$-compilation strategy of \citep{Algaba2024LowdepthSimulationsFermionicb} is tailored to high-weight transfer operators and relies on the fermionic-SWAP ($fSWAP$) gates. This strategy is particularly suitable for hardware platforms that implement the $fSWAP$ gate natively, but it yields reduced compilation efficiency when compiled on standard gate sets with single-qubit rotations and $CX$ entangling gates. Numerically, we obtain depth-8 (resp. depth-14) circuits for the even and odd horizontal (resp. vertical) hoppings.

\begin{figure}[ht!]
    \centering
    \includegraphics[width=0.8\linewidth]{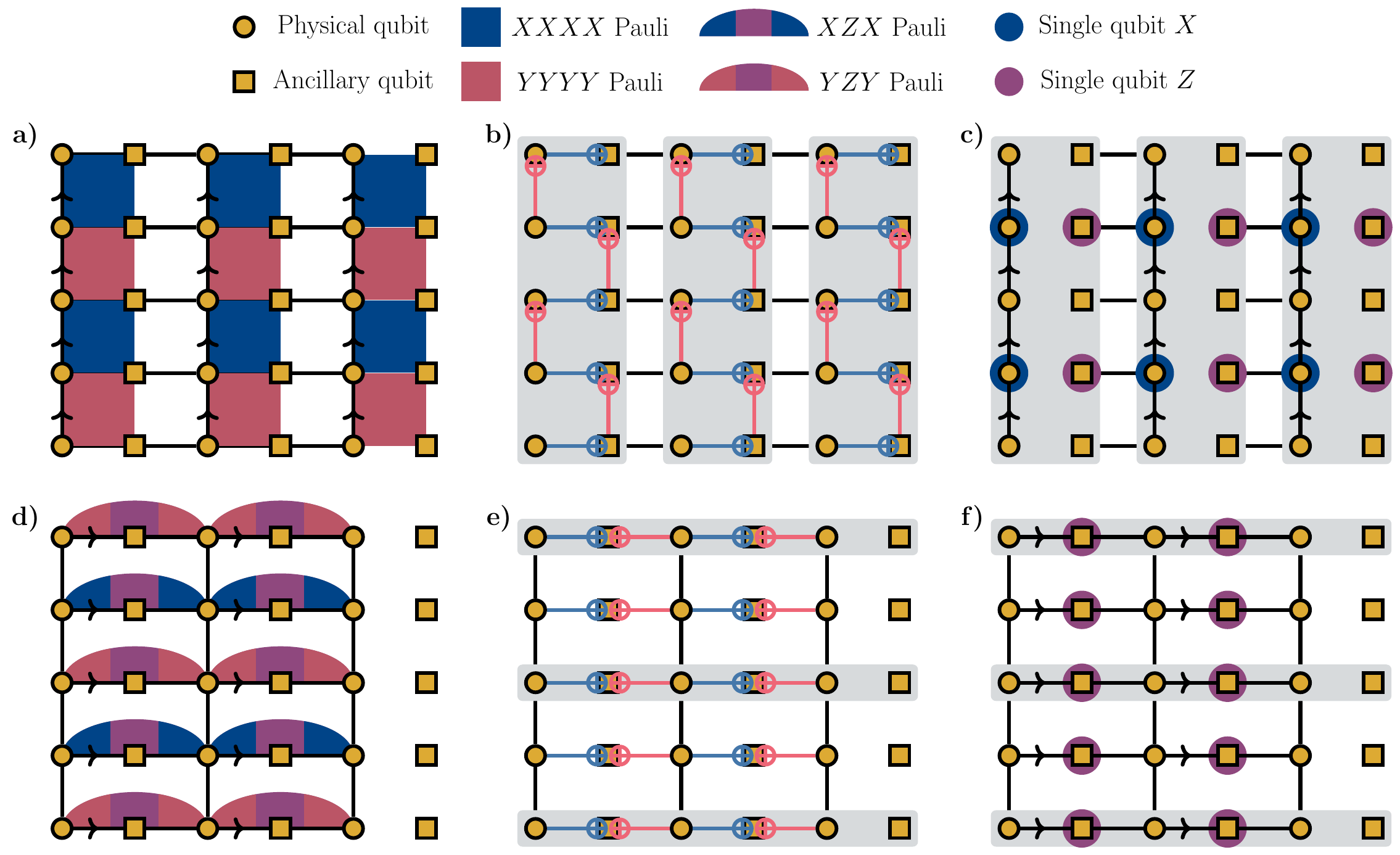}
        \caption{\textbf{a)} and \textbf{d)} Representation of the north-oriented and east-oriented flow sets for the transfer operators in the Verstraete-Cirac encoding. The circles represent the physical qubits and the squares represent the ancilla qubits. \textbf{b)} and \textbf{e)} Clifford encoding unitary circuits for the square-type (resp. triangular-type) vertical (resp. horizontal) flow sets. For convenience, single-qubit rotations are not explicitly included in the circuit. In particular, we adopt the convention of \cref{sec:encoding_unitaries_1D} with $XXXX/ZZZZ$ square stabilizers and $ZZZ$ triangle stabilizer. \textbf{c)} and \textbf{f)} After the Clifford encoding circuit, the twelve vertical stabilizers (resp. the ten horizontal stabilizers) are mapped to the twelve (resp. ten) single-qubit stabilizers $Z$, highlighted in blue, or destabilizers $X$, highlighted in red.}
    \label{fig:placeholder_vc_encoding}
\end{figure}

As an alternative to these standard compilation approaches, we propose to use the horizontal and vertical size-L flow set of \cref{fig:graph_coloring}. Indeed, the transfer operators in the Verstraete-Cirac encoding can be naturally split according to their Pauli weight, with square-shaped weight-4 vertical hoppings and triangular-shaped weight-3 horizontal hoppings represented graphically in \Cref{fig:placeholder_vc_encoding}. Additionally, the CCs in this case are one-dimensional lines of the interaction graph, which remain non-overlapping after the fermion-to-qubit encoding. This guarantees that all CC of each flow sets can be executed simultaneously. Furthermore, both triangular-shaped and square-shaped flow sets have optimal depth-2 unitary encoding circuit, whose circuits are reproduced in \Cref{fig:placeholder_vc_encoding} for convenience.

The Trotter unitary can be expressed as
\begin{align}\label{eq:vc_trotter_unitary}
    \exp(&-idt H) \stackrel{ST1}{\approx} \exp(-idt H_{NO})\exp(-idt H_{SO})\exp(-idt H_{WE})\exp(-iJdt H_{EA})\, , \\
    &=\exp(-idt H_{NO})\left(\prod_{l\in \text{phys.}} R^Z_l(\tfrac{\pi}{2}) \right)\exp(-idt H_{NO})\exp(-idt H_{EA})\left(\prod_{l\in \text{phys.}}R^Z_l(\tfrac{\pi}{2}) \right)^\dag \exp(-idt H_{EA})\, ,
\end{align}
where, going from the first to the second line, we used the additional identities between transfer operators detailed in \cref{eq:trotter_hdown,eq:trotter_heast}. The remaining evolution terms have already been derived in \cref{eq:vc_encoding_tij_tji_identity}, but we now explicitly include also all the layers of single-qubit rotations,
\begin{align}
    \exp(-idt \hat{H}_{NO}) & = \left(\prod_{l\in \text{phys.+anc.}}R_l^X(\tfrac{\pi}{2}) \right)\hat{U}_\text{square} \left(\prod_{(jk) \in F_\text{CC}} R^Z_{\mathrm{enc}(jk)}(-Jdt) \right)\hat{U}_\text{square}^\dag \left(\prod_{l\in \text{phys.+anc.}}R_l^X(\tfrac{\pi}{2}) \right)^\dag\, ,\\
    \exp(-idt \hat{H}_{EA}) & = \left(\prod_{l\in \text{phys.}}R_l^{X/Y}(\tfrac{\pi}{2}) \right)\hat{U}_\text{triangle} \left(\prod_{(j'k') \in F'_\text{CC}} R^Z_{\mathrm{enc}(j'k')}(-2dt) \right)\hat{U}_\text{triangle}^\dag \left(\prod_{l\in \text{phys.}}R_l^{X/Y}(\tfrac{\pi}{2}) \right)^\dag \, .
\end{align}
In the last line, the layer of single-qubit basis transformation $R_l^{X/Y}$ is $R^Y$ for even rows and $R^X$ for odd rows. 

To conclude our discussion, we provide conservative estimates of the circuit depth obtained with the petal flow sets and with our compilation strategy based on line flow sets. By naively compiling the petal flow sets, one implements even and odd rounds of depth-8 (resp. depth-12) unitaries, yielding an overall depth-40 unitary circuit. The optimization of \citep{Cade2020StrategiesSolvingFermiHubbarda} and \citep{Algaba2024LowdepthSimulationsFermionicb} can be applied for these same petal flow sets. However, this leads to limited improvements when assuming a quantum processor with a square-lattice nearest-neighbor connectivity and implementing a single entangling gate $CX$. For example, in the XYZ scheme~\citep{Algaba2024LowdepthSimulationsFermionicb}, the $fSWAP$ gate is compiled into 3 $CX$-layers, which would yield an overall circuit depth of $2\times (8+14) = 44$. In comparison, our compilation strategy using line flow sets yields a total unitary implementation with depth-16 (see \cref{eq:vc_trotter_unitary}), and this without accounting for any potential case-specific optimization.

\subsection{Detailed derivation for the Derby-Klassen encoding}
\label{sec:appendix_dk_encoding}

The Derby-Klassen encoding~\cite{Derby2021CompactFermionQubit} is a fermion-to-qubit encoding with a qubit-to-fermion ratio 3-to-2. As for the VC case, the individual fermionic parities are encoded as a weight-1 local operator. Following the original derivation, we split a square lattice into a bipartition of even and odd plaquettes. An ancillary degree of freedom is introduced at the center of each odd plaquette. Each edge $(j,k)$ is given an orientation (see Fig. 1. of the original manuscript, Ref.~\cite{Derby2021CompactFermionQubit}), and $f(j,k)$ represents the closest ancilla qubit to edge $(j,k)$ when it exists. The operator isomorphism is given, up to single-qubit rotations, by the map
\begin{align}
    \hat{V}_{j} &= Z_{j}\\
    \hat{E}_{jk} &= \begin{cases}
        (X_{j} Y_{k}) X_{f(j,k)} & \text{if $(j,k)$ oriented south}\, ,\\
        -(X_{j} Y_{k}) X_{f(j,k)} & \text{if $(j,k)$ oriented north}\, ,\\
        (X_{j} Y_{k}) Y_{f(j,k)} & \text{if $(j,k)$ horizontal}\, .
    \end{cases}
\end{align}
\begin{figure}[ht!]
    \centering
    \includegraphics[width=0.8\linewidth]{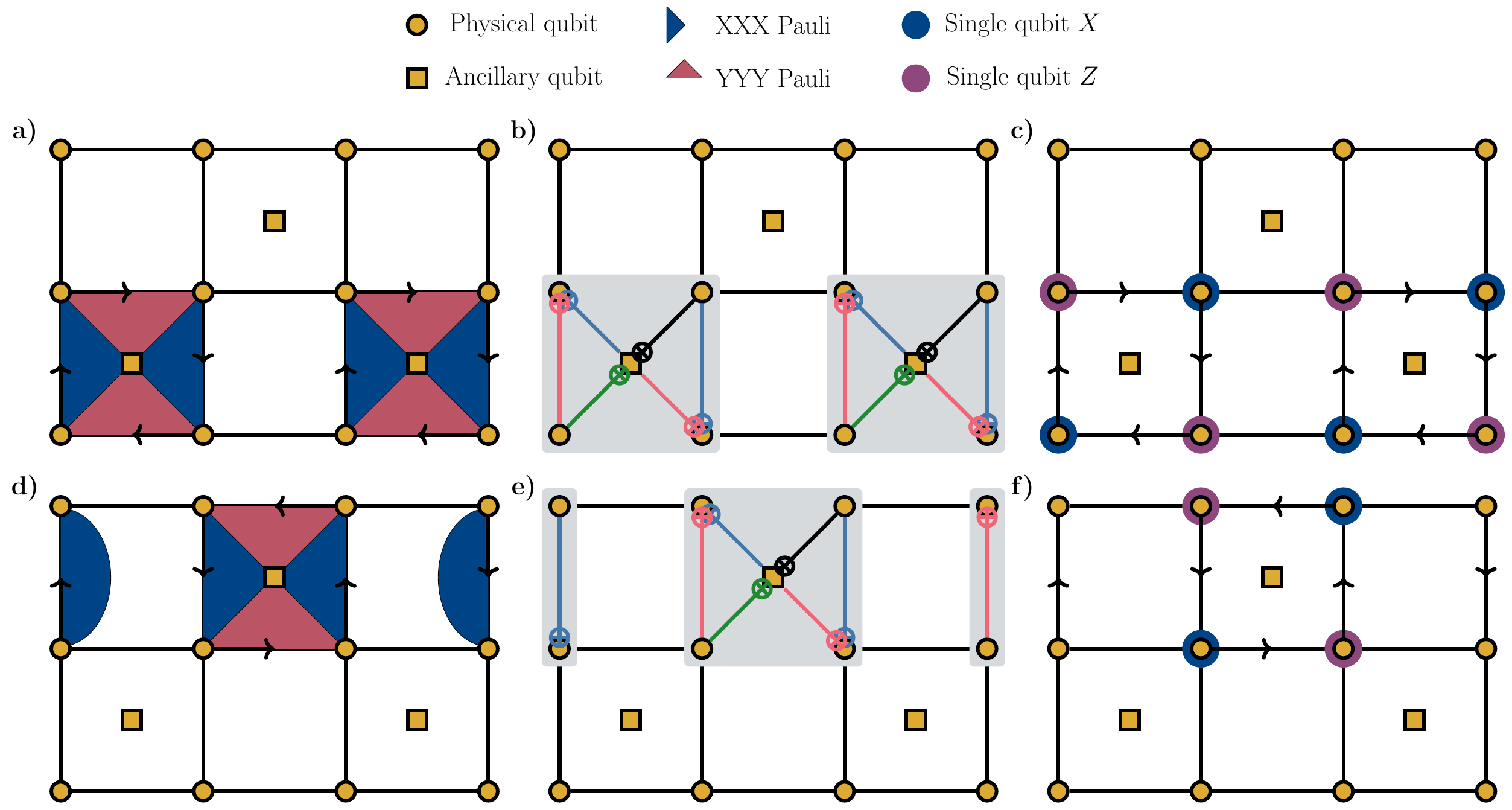}
    \caption{\textbf{a)d)} Representation of the \emph{even} and \emph{odd plaquette} flow sets for the transfer operators in the Derby-Klassen encoding. \textbf{b)e)} Special triangular-type Clifford encoding unitary circuits for a finite and periodic lattice. For convenience, the circuits are given without explicitly including single-qubit rotations and by assuming the convention of \cref{sec:encoding_unitaries_1D} with $XXX/ZZZ$ triangular stabilizers. \textbf{c)f)} Single qubit stabilizers $Z$ or destabilizers $X$ resulting from the application of the Clifford transformation $\hat{U}_\text{triangle,periodic}$ to the underconstrained stabilizer groups defined in a) and d).}    \label{fig:square_lattice_dk}
\end{figure}
The corresponding transfer operators are
\begin{align}
    \hat{T}_{jk} &= \begin{cases}
        (Y_{j} Y_{k}) X_{f(j,k)} & \text{if $(j,k)$ oriented south}\, ,\\
        -(Y_{j} Y_{k}) X_{f(j,k)} & \text{if $(j,k)$ oriented north}\, ,\\
        (Y_{j} Y_{k}) Y_{f(j,k)} & \text{if $(j,k)$ horizontal}\, .
    \end{cases}
\end{align}

For the Derby-Klassen encoded transfer operators, the horizontal and line flow sets are not a suitable flow-set decomposition. Indeed, two non-overlapping parallel lines of fermionic operators separated by a single unit cell are mapped to overlapping qubit operators (on the ancilla qubits). This prevents the corresponding Trotter terms to be applied in parallel. Instead, the plaquette flow sets defined in \cref{fig:graph_coloring}\textbf{b)} give the largest non-overlapping sets of commuting operators. These are represented in \cref{fig:square_lattice_dk} and correspond to the special triangular-shaped transfer operators for a size $L=4$ finite lattice with periodic boundary conditions. Notably, plaquette compilation was already discussed in \citep{Campbell2022EarlyFaulttolerantSimulations} as the optimal approach for square lattices, although not with the stabilizer formalism. Our approach does not yield a depth-optimal qubit stabilizer-encoding circuit, although further optimization may be possible within the plaquette flow set.

\section{Extension to 2D encodings with delocalized parities} \label{sec:appendix_parity_nonlocal_encodings}

In this section, we extend the discussion of \cref{sec:representation_fermionic_quadratic_operators} to describe one-dimensional qubit encodings where the individual fermionic parities $V_j$ are not represented by a single-qubit operator. In this case, we already note that the time-evolution unitary for the on-site interaction terms in the Fermi-Hubbard Hamiltonian, which we did not discuss in this work as it can be easily executed, becomes non-trivial. This is due to the fact that the on-site repulsion is local only when expressed in terms of the eigenbasis of the individual fermionic parities.

\subsection{Representation of fermionic quadratic operators on 1D lattices with non-overlapping delocalized parities}\label{sec:appendix_delocalized_non_overlapping}

For fermion-to-qubit encodings with qubit-to-fermion ratios 2-to-1, one can choose to delocalize the fermionic parity operator $V_j$ over two qubits, denoted as $v(j)$ and $h(j)$. Without loss of generality, we set $\hat{V}_j=Z_{v(j)} Z_{h(j)}$ for the rest of this discussion as all choices of two-qubit operators on $v(j)$ and $h(j)$ are equivalent up to single qubit rotations.

\begin{figure*}[ht!]
    \centering
    \includegraphics[width=0.8\linewidth]{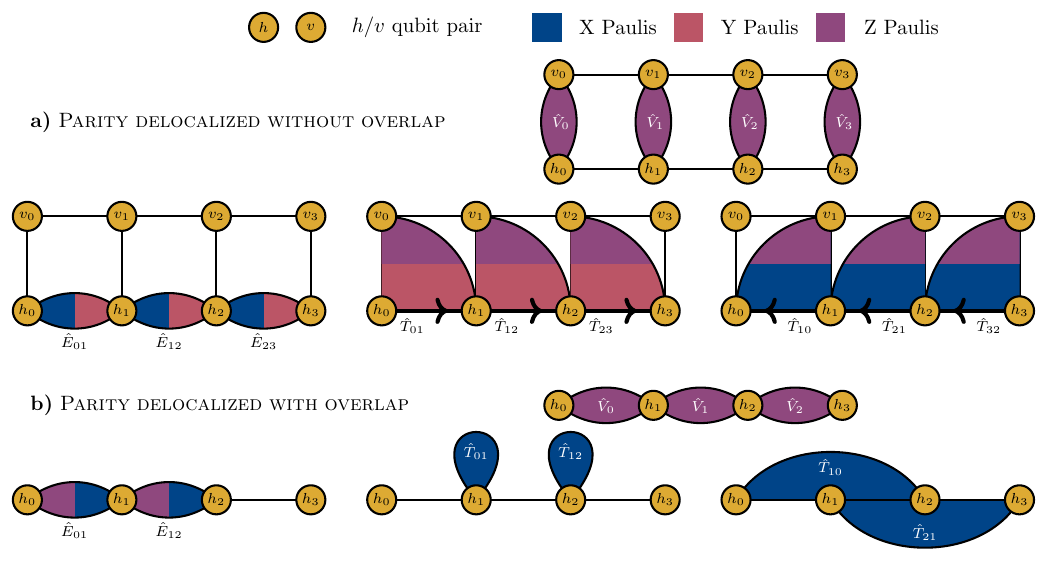}
    \caption{Edge and transfer operator representation for a one-dimensional chain of fermions with individual fermionic parities $V_i$ delocalized onto pairs of qubits. \textbf{a)} The parities can be delocalized over non-overlapping pairs of qubits $h(i)$ and $v(i)$. Example edge and transfer operators which recover the triangular-shaped scheme of \cref{fig:flavors_of_jw} with $h$ qubits playing the role of \emph{physical qubits}. Although this particular choice of edge operators purposefully breaks the symmetry between $h$ and $v$ qubits, this is not a requirement of delocalized encodings. \textbf{b)} The parities can be delocalized over overlapping pairs of qubits $h(i)$ and $h(i+1)$. A minimal set of edge and transfer operators satisfying the mixed-commutation relations has weight-1 and weight-3 transfer operators.}
    \label{fig:flavors_of_jw_delocalized}
\end{figure*}

In \cref{fig:flavors_of_jw_delocalized}, we show an example of the qubit representation of edge and transfer operators on a one-dimensional fermionic lattice using such a representation of the fermionic parity. In this example, the definition of the edge and transfer operators on the $h(j)$ and $v(j)$ qubits is not symmetric as in the previous cases. In fact, only $h(j)$ plays a role similar to the physical qubit in the previous constructions. We then find a direct correspondence with the \emph{right-oriented} triangular shaped transfer operators of~\cref{sec:representation_fermionic_quadratic_operators}, which implies that the Clifford encoding circuit depicted in~\cref{fig:clifford_circuit}a) can also be used for the parity-delocalized transfer operators in~\cref{fig:flavors_of_jw_delocalized}.

Because the local fermionic parities have been delocalized, one must derive a new qubit version of the product equality $\hat{T}_{jk} = - \hat{V}_j \hat{V}_k \hat{T}_{kj}$. With our definition of the transfer operators, we chose that $\hat{T}_{jk}$ have either $X$ or $Y$ on the $h$-qubits, and $Z$ on the $v$-qubits. One can then rewrite the multiplication on the left by $Z_{h(j)}Z_{v(j)}Z_{h(k)}Z_{v(k)}$ into the composition with two unitaries: one single-qubit basis change on the $h$-qubits and one global translation unitary on the $v$-qubits, denoted as $\mathcal{T}_{v}: v(i)\mapsto v(i+1)$
\begin{align}
    \hat{T}_{jk} &= - \hat{V}_j \hat{V}_k \hat{T}_{kj}\\
    &=-(Z_{v(j)} Z_{v(k)}) (Z_{h(j)} Z_{h(k)} \hat{T}_{kj})\\
    &= [\mathcal{T}_{v} R^Z_{h(j)}(\tfrac{\pi}{2}) R^Z_{h(k)}(\tfrac{\pi}{2}) ] \hat{T}_{kj} [R^Z_{h(j)}(\tfrac{\pi}{2}) R^Z_{h(k)}(\tfrac{\pi}{2}) \mathcal{T}_{v}]^\dag \label{eq:tjk_tkj_identity_delocalized_v2}
\end{align}
While the translation operator corresponds, in general, to a large-depth unitary transformation, we will show on the example of the Generalized Superfast encoding (GSE) that its implementation can almost entirely be avoided given certain arrangements of the $h/v$ qubits, see \cref{sec:appendix_gse_encoding}.

\subsection{Representation of fermionic quadratic operators on 1D lattices with overlapping delocalized parities}
It is also possible to construct representations of the fermionic operators where the qubit operator encoding the individual parities have overlapping support. Such a mapping yields time-evolution unitary of the kinetic part of the Hamiltonian with remarkable properties. The example we consider in \cref{fig:flavors_of_jw_delocalized}b requires $N_f+1$ qubits to encode $N_f$ fermions and is based on the following isomorphism
\begin{align}
    \hat{V}_{j}  &= Z_{j} Z_{j+1}\, , \quad\hat{E}_{j,j+1}= Z_{j} X_{j+1}\, ,\\
    \hat{T}_{j,j+1} &= X_{j+1}\, , \quad\hat{T}_{j+1,j} = Z_{j} X_{j+1} Z_{j+2}\, .
\end{align}
This example can be understood as a one-dimensional Kramers-Wannier dual~\citep{OBrien2025LocalJordanWignerTransformations,Lootens2024DualitiesOneDimensionalQuantum,Lootens2025LowDepthUnitaryQuantum,Shukla2020TensorNetworkApproach} to the bare JW transformation introduced in the main text and extends the classification of~\cref{sec:representation_fermionic_quadratic_operators}. It also appears in prior works on \emph{free-fermions in disguise}~\citep{SzaszSchagrin2025ConstructionSimulabilityQuantum,Fukai2025FreeFermionsDisguise,Fendley2019FreeFermionsDisguise,Fendley2024FreeFermionsJordan}. Our proposed representation has a west-oriented flow set with weight-1 transfer operators and an east-oriented flow set with weight-3 transfer operators. We can then show that the operator associated with the two flow sets are related by two layers of $CZ$ gates, on the even and odd edges respectively. Indeed,
\begin{equation}
    [CZ_{j,j+1} CZ_{j+2,j+1} ] X_{j+1} [CZ_{j,j+1} CZ_{j+2,j+1} ]^\dag = Z_j X_{j+1} Z_{j+2}
\end{equation}
It follows that the qubit representation of the fermionic unitary representation of the west-oriented flow set has depth-0 (\textit{i.e.}, it is composed solely of single-qubit gates), while that of east-oriented flow set has depth-4, which are both depth-optimal for weight-1 Paulis and weight-3 Paulis respectively.

This qubit representation then combines a depth-optimal circuit decomposition of the time-evolution unitary as well as a small qubit-to-fermion ratio $1 + 1/N_f$. While this appears to contradict our observation of \emph{space-time} tradeoffs in fermion-to-qubit encodings, this is a direct consequence of the delocalization of the individual fermionic parities. A circuit representation of the Fermi-Hubbard Hamiltonian that includes both a kinetic component as well as an on-site repulsion component will require \emph{relocalizing} the individual fermionic parities when implementing the time-evolution operator for the interaction term. The idea of a \emph{space-time} tradeoff then requires including both kinetic and interaction part of the Hamiltonian. This can be linked to the discussion of~\citep{Cade2020StrategiesSolvingFermiHubbarda,Jiang2018QuantumAlgorithmsSimulatea}, where the 1D Fast-Fourier transformation is reformulated as a transformation to the parity basis.

\subsection{Application to the generalized superfast encoding}\label{sec:appendix_gse_encoding}
\begin{figure}[ht!]
    \centering
    \includegraphics[width=\linewidth, page=1]{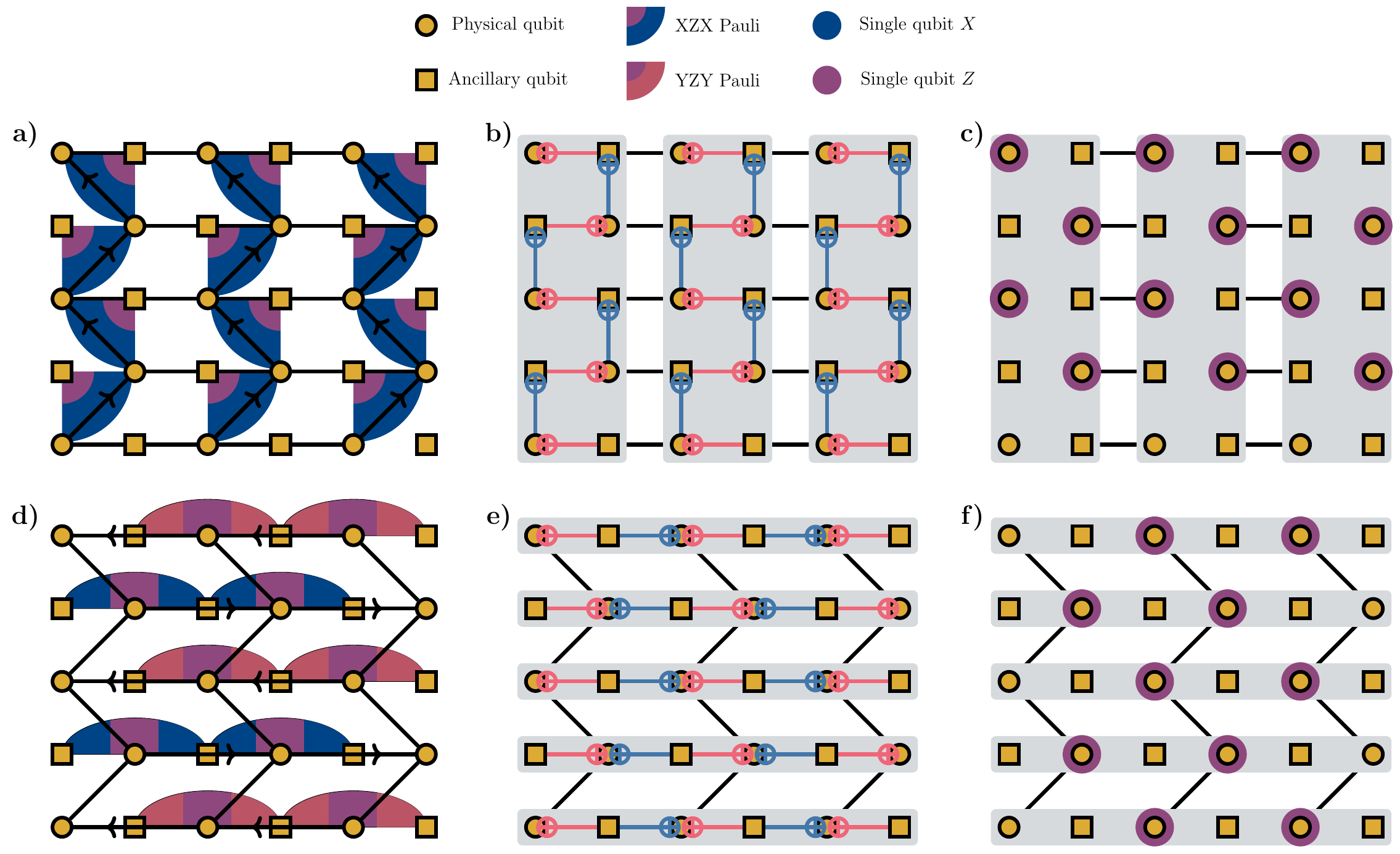}
    \caption{Characterization of two out of the four flow sets in the GSE mapping. \textbf{a)} Representation of the triangular-type north-oriented flow sets. \textbf{b)}\textbf{c)} Triangular-type encoding circuit of \cref{fig:clifford_circuit} and corresponding location of single-qubit transformed $Z$ stabilizers. \textbf{d)} Alternating east- and west- oriented flow sets are also of the triangular type. \textbf{e)f)} Encoding circuits and transformed stabilizers location for the alternating horizontal flow set.}
    \label{fig:placeholder_superfast_operators}
\end{figure}

\begin{figure}[ht!]
    \centering
    \includegraphics[width=\linewidth, page=2]{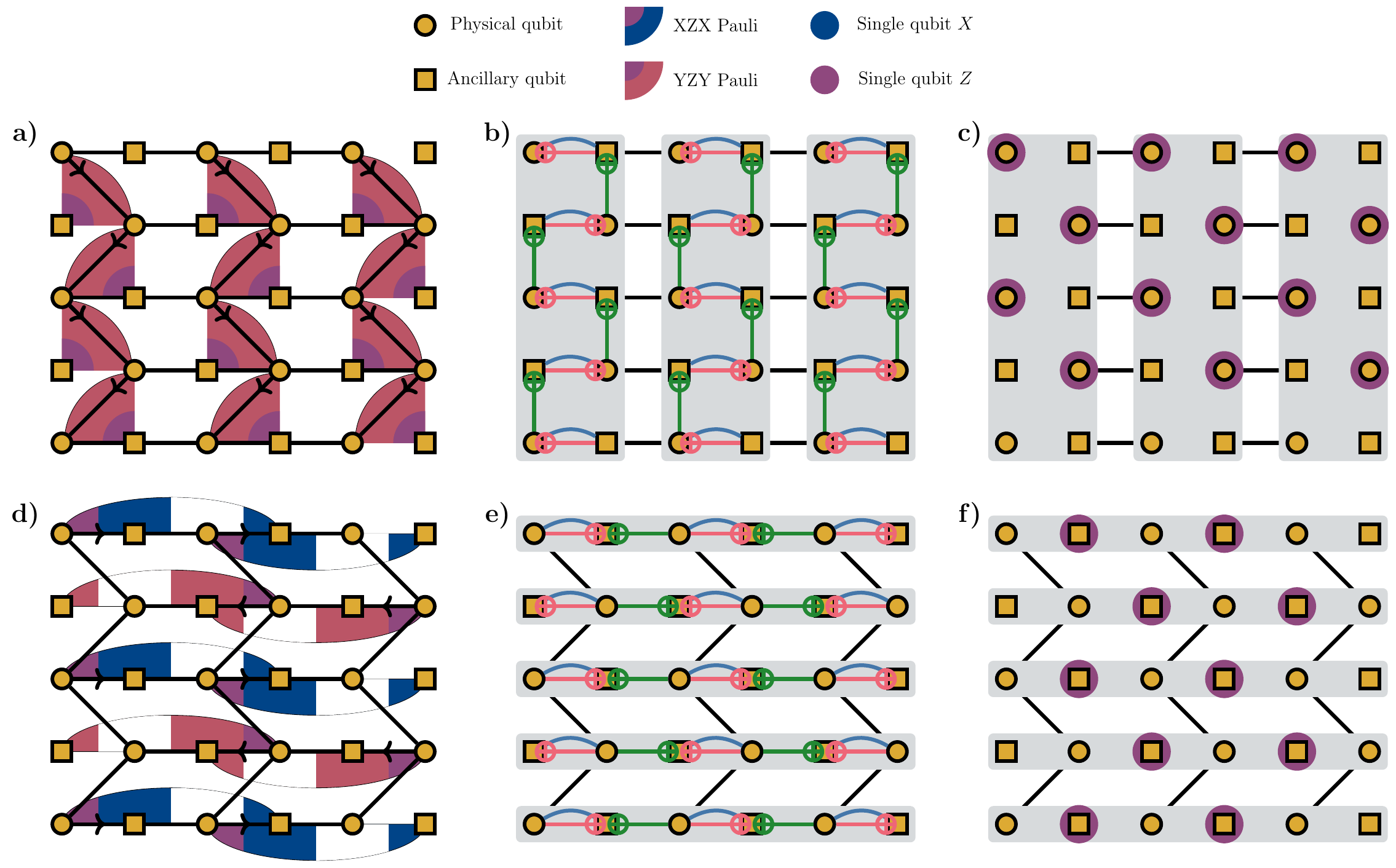}
    \caption{Characterization of two remaining flow sets in the GSE mapping. \textbf{a)} The south-oriented flow set are also of the triangular type, but on a different set of qubits. \textbf{b)}\textbf{c)} Symmetric layout of the triangular-type encoding circuit of \cref{fig:clifford_circuit} and corresponding location of single-qubit transformed $Z$ stabilizers. \textbf{d)} Alternating west- and east- oriented flow sets are brought to a triangular type form by a single layer of horizontal $SWAP$ exchange between the $h$-qubits and the $v$-qubits. \textbf{e)f)} The corresponding encoding circuits is the composition of a $SWAP$-layer and a translated triangular-type encoding sequence.}
    \label{fig:placeholder_superfast_operators_v2}
\end{figure}

The generalized superfast encoding (GSE)~\citep{Setia2019SuperfastEncodingsFermionic} is a fermion-to-qubit encoding with qubit-to-fermion ratio 1-to-2 and with the fermionic parity encoded as a weight-2 Pauli operator. While this setup deviates slightly from the ones discussed in the main text, we show here that the key messages of our work apply also to this case with minor changes. Each fermionic site $i$ will be associated with two qubit indices $v(i)$ and $h(i)$. However, there is no longer a \emph{physical index} because the parity is delocalized over qubits $v(i)$ and $h(i)$. The GSE encoded vertex- and edge- and transfer- operators are defined as
\begin{align}
    \hat{V}_{j} &= Z_{v(j)} Z_{h(j)}\\
    \hat{E}_{jk} &= \begin{cases}
        (X_{v(j)} Y_{v(k)}) Z_{h(j)} Z_{h(k)} & \text{if $(jk)$ is vertical and north-oriented}\, ,\\
        (X_{h(j)} Y_{h(k)}) & \text{if $(jk)$ is horizontal and east-oriented}\, .
    \end{cases}\\
    \hat{T}_{jk} &= \begin{cases}
        (Y_{v(j)} Y_{v(k)}) Z_{h(k)} & \text{if $(jk)$ is vertical and north-oriented}\, ,\\
        (Y_{h(j)} Y_{h(k)}) Z_{v(j)} & \text{if $(jk)$ is horizontal and east-oriented\, ,}
    \end{cases}
\end{align}
also depicted in \Cref{fig:placeholder_superfast_operators} for qubits placed on a square lattice topology.
The encoding circuits for vertical and horizontal hoppings are the same as the triangular-shaped Clifford encoding of \cref{fig:clifford_circuit} and are repeated in \Cref{fig:placeholder_superfast_operators}. One key difference is that now the local fermionic parities have been delocalized as in \cref{sec:appendix_delocalized_non_overlapping} and one must use the corresponding product equality when encoding the transfer operators
\begin{align}
    \hat{T}_{jk} &= - \hat{V}_j \hat{V}_k \hat{T}_{kj} = - Z_{h(j)} Z_{v(j)} Z_{h(k)} Z_{v(k)} \hat{T}_{kj}\\
    & = \begin{cases}
       (X_{v(j)} X_{v(k)}) Z_{h(j)} & \text{if $\langle jk\rangle$ is vertical}\\
       (X_{h(j)} X_{h(k)}) Z_{v(k)} & \text{if $\langle jk\rangle$ is horizontal}
    \end{cases}\\
    & = \begin{cases}
       [\mathcal{T}_h R^Z_{v(j)}(\tfrac{\pi}{2}) R^Z_{v(k)}(\tfrac{\pi}{2})] \hat{T}_{kj} [\mathcal{T}_h R^Z_{v(j)}(\tfrac{\pi}{2}) R^Z_{v(k)}(\tfrac{\pi}{2})]^\dag& \text{if $\langle jk\rangle$ is vertical}\\
       [\mathcal{T}_v R^Z_{h(j)}(\tfrac{\pi}{2}) R^Z_{h(k)}(\tfrac{\pi}{2})] \hat{T}_{kj} [\mathcal{T}_v R^Z_{h(j)}(\tfrac{\pi}{2}) R^Z_{h(k)}(\tfrac{\pi}{2})]^\dag& \text{if $\langle jk\rangle$ is horizontal}
    \end{cases}
\end{align}
For the vertical operators, we find that a symmetric version of the Clifford encoding circuit for the north-oriented flow sets also works for the south-oriented flow sets, see \cref{fig:placeholder_superfast_operators_v2}\textbf{a}. For the horizontal hoppings, we find after a few simplifications that we must additionally implement a single layer of $SWAP$ between all the qubit indices $h(i), v(i)$ as well as layers $R^Z_{h/v}$ of single-qubit $R^Z(\tfrac{\pi}{2})$ gate on the $h/v$ qubits, see \cref{fig:placeholder_superfast_operators_v2}\textbf{d}. The encoding circuits for the west-oriented and south-oriented flow sets are depicted in \cref{fig:placeholder_superfast_operators_v2}.
Lastly, the Trotter unitary can be expressed as a depth-16 unitary with 2 additional nearest-neighbor SWAP layers
\begin{align}
    &\exp(-idt H)\\
    &= \exp(-idt H_{NO})\exp(-idt H_{SO})\exp(-idt H_{EA})\exp(-idt H_{WE}) \\
    &=\exp(-idt H_{NO}) (R^Z_v) (h\leftrightarrow v)\exp(-idt H_{NO}) (R^Z_h R^Z_v )\exp(-idt H_{EA})(h\leftrightarrow v) (R^Z_h) \exp(-idt H_{EA})
\end{align}

\end{document}